\documentclass{article}%
\usepackage{hyperref}
\usepackage{amsmath}
\usepackage{amsfonts}
\usepackage{amssymb}
\usepackage{graphicx}%
\begin{document}

\title{Field-dependent BRST-antiBRST Transformations in Yang--Mills and
Gribov--Zwanziger Theories}
\author{\textsc{Pavel Yu. Moshin${}^{a}$\thanks{moshin@rambler.ru \hspace{0.5cm}
${}^{\dagger}$reshet@ispms.tsc.ru}\ \ and Alexander A. Reshetnyak${}%
^{b,c\dagger}$}\\\textit{${}^{a}$National Research Tomsk State University, 634050, Russia,}\\\textit{${}^{b}$Institute of Strength Physics and Materials Science}\\\textit{Siberian Branch of Russian Academy of Sciences, 634021, Tomsk,
Russia,}\\\textit{${}^{c}$Tomsk State Pedagogical University, 634061, Russia }}
\maketitle

\begin{abstract}
We introduce the notion of finite BRST-antiBRST transformations, both global
and field-dependent, with a doublet $\lambda_{a}$, $a=1,2$, of anticommuting
Grassmann parameters and find explicit Jacobians corresponding to these
changes of variables in Yang--Mills theories. It turns out that the finite
transformations are quadratic in their parameters. At the same time, exactly
as in the case of finite field-dependent BRST transformations for the
Yang--Mills vacuum functional, special field-dependent BRST-antiBRST
transformations, with $s_{a}$-potential parameters $\lambda_{a}=s_{a}\Lambda$
induced by a finite even-valued functional $\Lambda$ and by the anticommuting
generators $s_{a}$ of BRST-antiBRST transformations, amount to a precise
change of the gauge-fixing functional. This proves the independence of the
vacuum functional under such BRST-antiBRST transformations. We present the
form of transformation parameters that generates a change of the gauge in the
path integral and evaluate it explicitly for connecting two arbitrary $R_{\xi
}$-like gauges. For arbitrary differentiable gauges, the finite
field-dependent BRST-antiBRST transformations are used to generalize the
Gribov horizon functional $h$, given in the Landau gauge, and being an
additive extension of the Yang--Mills action by the Gribov horizon functional
in the Gribov--Zwanziger model. This generalization is achieved in a manner
consistent with the study of gauge independence. We also discuss an extension
of finite BRST-antiBRST\ transformations to the case of general gauge theories
and present an ansatz for such transformations.

\end{abstract}

\vspace{12mm} \noindent\textsl{Keywords:} \ BRST-antiBRST Lagrangian
quantization, gauge theories, Yang--Mills theory, Gribov--Zwanziger theory,
field-dependent BRST-antiBRST transformations

\section{Introduction}

Contemporary quantization methods for gauge theories
\cite{books1,books2,books3,books4} are based primarily on the special
supersymmetries known as BRST symmetry \cite{BRST1,BRST2,BRST3} and
BRST-antiBRST symmetry \cite{aBRST1,aBRST2,aBRST3,aBRST4}. They are
characterized by the presence of a Grassmann-odd parameter $\mu$ and two
Grassmann-odd parameters $(\mu,\bar{\mu})$, respectively. In the framework of
the $\mathrm{Sp}\left(  2\right)  $-covariant schemes of generalized
Hamiltonian \cite{BLT1h,BLT2h} and Lagrangian \cite{BLT1,BLT2} quantization
(see also \cite{GH1,Hull}), the parameters $(\mu,\bar{\mu})\equiv(\mu_{1}%
,\mu_{2})=\mu_{a}$ form an $\mathrm{Sp}\left(  2\right)  $-doublet. These
infinitesimal odd-valued parameters may be regarded as constants and thus used
to derive the Ward identities. They may also be chosen as field-dependent
functionals and thus used to establish the gauge-independence of the
corresponding vacuum functional in the path integral approach.

BRST transformations with a finite field-dependent parameter in Yang--Mills
theories, whose quantum action is constructed by the Faddeev--Popov rules
\cite{FP}, were first introduced in \cite{JM} by means of a functional
equation for the parameter in question, and used to provide the path integral
with such a change of variables that would allow one to relate the quantum
action in a certain gauge with the one given in a different gauge; see also
\cite{JM1}. This equation, as well as a similar equation \cite{RM} for the
finite parameter of a field-dependent BRST transformation in the generalized
Hamiltonian formalism, has not been solved in the general setting of the
problem. Namely, the corresponding equation (4.13) in \cite{JM}, or equation
(3.6) in \cite{RM}, for the Jacobian $J$ of a change of variables given by
infinitesimal field-dependent BRST transformations with an odd-valued
functional\footnote{$\Theta^{\prime}(\phi(\kappa))$ depends on a numerical
parameter, $\kappa$, so that the finite field-dependent BRST transformations
with the odd-valued functional $\Theta(\phi(0))$ are obtained from
$\Theta^{\prime}(\phi(\kappa))$ by $\Theta(\phi(0)) = \int\limits^{1}_{0}
\Theta^{\prime}(\phi(\kappa)) dk$.} $\Theta^{\prime}(\phi(\kappa))$ allows one
to express an additional contribution $S_{1}$ to the quantum action in terms
of $\Theta(\phi(0))$, but has not been solved neither in the form $S_{1}%
=S_{1}(\Theta(\phi(0)))$, for an unknown quantity $S_{1}$, nor in
the form $S_{1}=S_{1}(\Theta(\phi(0)))$, for an unknown quantity
quantity $\Theta (\phi(0))$. Instead, a series of particular cases
having the form of an ansatz for the functional $S_{1}$ have been
examined, and a solution of the above-mentioned equation was found
without any explicit calculation of the Jacobian for the change of
variables induced by finite field-dependent BRST
transformations.\footnote{The property of gauge independence for
the vacuum functional in the Yang--Mills theory with an action
constructed by the Faddeev-Popov recipe \cite{FP}, or with an
action constructed by the Batalin--Vilkovisky (BV) procedure
\cite{BV}, uses an explicit form of the above Jacobian.} On the
other hand, there emerges the problem of establishing a relation
of the Faddeev--Popov action in a certain gauge with the action in
a different gauge, by using a change of variables induced by a
finite field-dependent BRST transformation. This problem was
solved for the first time in the case of linear and quadratic
gauges in \cite{JM} and for the class of general gauges in
\cite{LL1}, thereby providing an exact relation between a finite
parameter and a finite variation of the gauge-fixing condition in
terms of the gauge Fermion. There it was established that the
Jacobian of any finite field-dependent BRST transformation
reproduces BRST-exact terms, which can be entirely absorbed into
the gauge-fixed part the of BRST-invariant Faddeev--Popov action,
corresponding to a certain change of the gauge $\Delta\psi$, so
that the vacuum functional $Z_{\psi+\Delta\psi}$, resulting from
the above change of variables, coincides with the initial vacuum
functional $Z_{\psi}$ and should be regarded as a vacuum
functional with the same BRST-exact classical (renormalized)
action, having, however, a gauge-fixed (BRST-exact) action given
by a different gauge, $\psi+\Delta\psi$. In particular, this
implies the conservation of the number of physical degrees of
freedom in a given Yang--Mills theory with respect to finite
field-dependent BRST transformations. This means the impossibility
of relating the Yang--Mills theory to a theory whose action may
contain, in addition to the Faddeev--Popov action, some BRST
non-invariant terms (such as the Gribov horizon functional in the
Gribov--Zwanziger theory \cite{Zwanziger}, having additional
degrees of freedom as compared to the Yang--Mills theory) in the
same configuration space.\footnote{Instead of a local
Gribov--Zwanziger horizon functional $S_{\gamma}$ in (3.3), there
exists a relation \cite{Upadhyay2} by finite field-dependent BRST
transformations to a BRST-invariant model with the functional
$\Sigma_{\gamma}$ in (3.6), being a Yang--Mills theory defined in
an appropriate configuration space.}

The solution of a similar problem for arbitrary dynamical systems with
first-class constraints in the generalized Hamiltonian formalism
\cite{BRST3,BFV,Henneaux1} has been recently proposed in \cite{BLThf}. For
general gauge theories, which may possess a reducible gauge symmetry and/or an
open gauge algebra, an exact Jacobian corresponding to a change of variables
given by field-dependent BRST transformations in the path integral constructed
according to the Batalin--Vilkovisky (BV) procedure \cite{BV} was obtained in
\cite{Reshetnyak} and shown to be identical with the Jacobian of the
Yang--Mills theory. The study of \cite{Reshetnyak} extends the results of
\cite{LL1} to first-rank theories with a closed algebra and solves the problem
of gauge-independence for gauge theories with the so-called soft breaking of
BRST symmetry. This problem was raised in \cite{llr1} to study the problem of
Gribov copies \cite{Gribov} by using various gauges in the Gribov--Zwanziger
approach \cite{Zwanziger}; for recent progress, see
\cite{sorellas,sorellas2,LL2,Gongyo,Reshetnyak2}.

On the other hand, there emerges the problem of finding a
correspondence of the quantum action in the BRST-antiBRST
invariant Lagrangian quantization \cite{BLT1,BLT2,BLT3}, where
gauge is introduced by a Bosonic gauge-fixing functional, $F$,
with the quantum action of the same theory in a different gauge,
$F+\Delta F$, for a finite value $\Delta F$, by using a change of
variables in the vacuum functional. This problem has not been
solved even in theories of Yang--Mills type. Note that finite
field-dependent antiBRST transformations in Yang--Mills theories
were considered in \cite{Upadhyay1} in the same way as in the case
of BRST transformations \cite{JM}, so as to relate the antiBRST
invariant quantum action of a Yang--Mills theory in different
gauges by using an ansats for a term introduced to the quantum
action in order to satisfy an infinitesimal functional equation
for the transformation parameter. The study of \cite{Upadhyay3}
proposed finite two-parametric BRST-antiBRST transformations
(``mixed'', by the terminology \cite{Upadhyay3}): ``$\delta_m \phi
= \overleftarrow{s}_\mathrm{a} \Theta_1 +
\overleftarrow{s}_{\mathrm{ab}} \Theta_2$" in (3.7), including
field-dependent ones, which form a Lie superalgebra; however,
without any parameters, constant and/or field-dependent, being
quadratic in $\Theta_1$, $\Theta_2$ (allowing one to consider
BRST-antiBRST transformations as group transformations), which
prohibits the complete BRST-antiBRST invariance of the quantum
action in Yang--Mills theories and similarly in more general gauge
theories. Therefore, this leads immediately to the problem of
finding a solution for the above functional equation, since the
latter does not ``feel'' the finite polynomial character of the
parameters $\Theta_{1}\cdot\Theta_{2}$, and therefore prohibits
the gauge independence of the vacuum functional under finite
field-dependent BRST-antiBRST transformations even for
functionally-dependent parameters (see footnote \ref{4}).

A similar problem in the $\mathrm{Sp}\left(  2\right)  $-covariant
generalized Hamiltonian formalism \cite{BLT1h,BLT2h} remains
unsolved\footnote{For the recent progress achieved in this area
since the appearance of the present work in arXiv, see footnote
\ref{HamMR} in Discussion.} as well. We expect that the solution
of these problems in the Lagrangian and Hamiltonian quantization
schemes for gauge theories should be based on the concept of
finite BRST-antiBRST transformations with an $\mathrm{Sp}\left(
2\right) $-doublet of Grassmann-odd parameters $\mu_{a}\left(
\phi\right) $ depending on the field variables. This would allow
one to generate the Gribov horizon functional by using different
gauges in a way consistent with the gauge-independence of the path
integral, based on the Gribov--Zwanziger prescription
\cite{Zwanziger} and starting from the BRST-antiBRST invariant
Yang--Mills quantum action in the Landau gauge.

Motivated by these reasons, we intend to address the following issues, paying
our attention primarily to the Yang--Mills theory in Lagrangian formalism:

\begin{enumerate}
\item introduction of \emph{finite BRST-antiBRST transformations}, being
polynomial in powers of a constant $\mathrm{Sp}\left(  2\right)  $-doublet of
Grassmann-odd parameters $\lambda_{a}$ and leaving the quantum action of the
Yang--Mills theory invariant to all orders in $\lambda_{a}$;

\item definition of \emph{finite field-dependent BRST-antiBRST
transformations}, being polynomial in powers of an $\mathrm{Sp}\left(
2\right)  $-doublet of Grassmann-odd functionals $\lambda_{a}(\phi)$ depending
on the classical Yang--Mills fields, the ghost-antighost fields, and the
Nakanishi--Lautrup fields; calculation of the Jacobian related to a change of
variables by using a special class of such transformations with $s_{a}
$-potential parameters $\lambda_{a}(\phi)=s_{a}\Lambda(\phi)$ for a
Grassmann-even functional $\Lambda(\phi)$ and Grassmann-odd generators $s_{a}
$ of BRST-antiBRST transformations;

\item solution of the so-called compensation equation for an unknown
functional $\Lambda$ generating the $\mathrm{Sp}\left(  2\right)  $-doublet
$\lambda_{a}$ with the purpose of establishing a relation of the Yang--Mills
quantum action $S_{F}$ in a gauge determined by a gauge Boson $F$ with the
quantum action $S_{F+\Delta F}$ in a different gauge $F+\Delta F$;

\item explicit construction of the parameters $\lambda_{a}$ of finite
field-dependent BRST-antiBRST transformations generating a change of the gauge
in the path integral within a class of linear $R_{\xi}$-like gauges realized
in terms of Bosonic gauge functionals $F_{\left(  \xi\right)  }$, with
$\xi=0,1$ corresponding to the Landau and Feynman (covariant) gauges, respectively;

\item construction of the Gribov horizon functional $h_{\xi}$ in arbitrary
$R_{\xi}$-like gauges by means of finite field-dependent BRST-antiBRST
transformations starting from a known BRST-antiBRST non-invariant functional
$h$, given in the Landau gauge and realized in terms of the Bosonic functional
$F_{\left(  0\right)  }$.
\end{enumerate}

The present work is organized as follows. In Section~\ref{gensetup}, we remind
the general setup of the BRST-antiBRST Lagrangian quantization of general
gauge theories and list its basics ingredients. In Section~\ref{fBRSTa}, we
introduce the notion of finite BRST-antiBRST transformations, both global and
local (field-dependent). We find an explicit Jacobian corresponding to this
change of variables in theories of Yang--Mills type and show that, exactly as
in the case of field-dependent BRST transformations for the Yang--Mills vacuum
functional \cite{LL1}, the field-dependent transformations amount to a precise
change of the gauge-fixing functional. In Section~\ref{YMgauges}, we present
the form of transformation parameters that generates a change of the gauge and
evaluate it for connecting two arbitrary $R_{\xi}$-like gauges in Yang--Mills
theories. In Section~\ref{GZ}, the Gribov horizon functional in an arbitrary
$R_{\xi}$-like gauge, and generally in any differentiable gauge, is determined
with the help of respective finite field-dependent BRST-antiBRST
transformations. In Discussion, we make an overview of our results and outline
some open problems. In particular, we discuss an extension of finite
BRST-antiBRST transformations to the case of general gauge theories and
present an ansatz for such transformations. In Appendix~\ref{AppC}, we study
the group properties of finite field-dependent BRST-antiBRST transformations.
In Appendix~\ref{AppA}, we present a detailed calculation of the Jacobian
corresponding to the finite, both global (Appendix~\ref{AppA1}) and
field-dependent (Appendix~\ref{AppA2}), BRST-antiBRST transformations.
Appendix~\ref{AppB} is devoted to calculations involving the BRST-antiBRST
invariant Yang--Mills action in $R_{\xi}$-gauges.

We use DeWitt's condensed notations \cite{DeW}. By default, derivatives with
respect to the fields are taken from the right, and those with respect to the
corresponding antifields are taken from the left; otherwise, left-hand and
right-hand derivatives are labelled by the subscripts ``$l$'' and ``$r$'',
respectively; $F,_{A}$ stands for the right-hand derivative $\delta
F/\delta\phi^{A}$ of a functional $F=F\left(  \phi\right)  $ with respect to
$\phi^{A}$. The raising and lowering of $\mathrm{Sp}\left(  2\right)  $
indices, $s^{a}=\varepsilon^{ab}s_{b}$, $s_{a}=\varepsilon_{ab}s^{b}$, is
carried out with the help of a constant antisymmetric second-rank tensor
$\varepsilon^{ab}$, $\varepsilon^{ac}\varepsilon_{cb}=\delta_{b}^{a}$, subject
to the normalization condition $\varepsilon^{12}=1$. The Grassmann parity and
ghost number of a quantity $A$, assumed to be homogeneous with respect to
these characteristics, are denoted by $\varepsilon\left(  A\right)  $,
$\mathrm{gh}(A)$, respectively.

\section{General Setup for BRST-antiBRST Lagrangian Quantization}

\label{gensetup}%
\renewcommand{\theequation}{\arabic{section}.\arabic{equation}} \setcounter{equation}{0}

The BRST-antiBRST Lagrangian quantization of general gauge theories
\cite{BLT1,BLT2,BLT3} involves a set of fields $\phi^{A}$ and a set of
corresponding antifields $\phi_{Aa}^{\ast}$ ($a=1,2$), $\bar{\phi}_{A}$, where
the doublets of antifields $\phi_{Aa}^{\ast}$ play the role of sources to the
BRST and antiBRST transformations, while the antifields $\bar{\phi}_{A}$ are
the sources to the mixed BRST and antiBRST transformations, with the following
distributions of the Grassmann parity and ghost number:%
\begin{equation}
\varepsilon(\phi^{A})\equiv\varepsilon_{A}\ ,\quad\varepsilon(\phi_{Aa}^{\ast
})=\varepsilon_{A}+1\ ,\quad\varepsilon(\bar{\phi}_{A})=\varepsilon
_{A}\ ,\quad\mathrm{gh}(\phi_{Aa}^{\ast})=(-1)^{a}-\mathrm{gh}(\phi
^{A})\ ,\quad\mathrm{gh}(\bar{\phi}_{A})=-\mathrm{gh}(\phi^{A})\ .
\label{antif}%
\end{equation}
The configuration space of fields $\phi^{A}$ is identical with that of the BV
formalism \cite{BV} of covariant quantization and is determined by the
properties of the initial classical theory. Namely, we consider an initial
classical theory of fields $A^{i}$, $\varepsilon(A^{i})\equiv\varepsilon_{i}$,
with an action $S_{0}(A)$ invariant under gauge transformations,
\begin{equation}
\delta A^{i}=R_{\alpha_{0}}^{i}(A)\zeta^{\alpha_{0}}\Longrightarrow
S_{0,i}(A)R_{\alpha_{0}}^{i}(A)=0\ , \label{Nid}%
\end{equation}
where $R_{\alpha_{0}}^{i}(A)$ are generators of the gauge transformations,
$\varepsilon(R_{\alpha_{0}}^{i})=\varepsilon_{i}+\varepsilon_{\alpha_{0}}$,
and $\zeta^{\alpha_{0}}$ are arbitrary functions of the space-time
coordinates, $\varepsilon(\zeta^{\alpha_{0}})=\varepsilon_{\alpha_{0}}$. The
generators $R_{\alpha_{0}}^{i}(A)$ form a gauge algebra \cite{BV} with the
relations%
\begin{align}
&  R_{\alpha_{0},j}^{i}(A)R_{\beta_{0}}^{j}(A)-\left(  -1\right)
^{\varepsilon_{\alpha_{0}}\varepsilon_{\beta_{0}}}R_{\beta_{0},j}%
^{i}(A)R_{\alpha_{0}}^{j}(A)=-R_{\gamma_{0}}^{i}(A)F_{\alpha_{0}\beta_{0}%
}^{\gamma_{0}}\left(  A\right)  -S_{0,j}(A)M_{\alpha_{0}\beta_{0}}^{ij}\left(
A\right)  \ ,\nonumber\\
&  F_{\alpha_{0}\beta_{0}}^{\gamma_{0}}=-\left(  -1\right)  ^{\varepsilon
_{\alpha_{0}}\varepsilon_{\beta_{0}}}F_{\beta_{0}\alpha_{0}}^{\gamma_{0}%
}\ ,\ \ M_{\alpha_{0}\beta_{0}}^{ij}=-\left(  -1\right)  ^{\varepsilon
_{i}\varepsilon_{j}}M_{\alpha_{0}\beta_{0}}^{ji}=-\left(  -1\right)
^{\varepsilon_{\alpha_{0}}\varepsilon_{\beta_{0}}}M_{\beta_{0}\alpha_{0}}%
^{ij}\,. \label{gauge_alg}%
\end{align}
In case the vectors $R_{\alpha_{0}}^{i}(A)$, enumerated by the index
$\alpha_{0}$, are linearly independent, the theory is irreducible; otherwise
it is reducible. Depending on the (ir)reducibility of the generators of gauge
transformations, the specific structure of the configuration space $\phi^{A}$
is described by the set of fields%
\begin{equation}
\phi^{A}=(A^{i},B^{\alpha_{s}|a_{1}...a_{s}},C^{\alpha_{s}|a_{0}...a_{s}%
})\ ,\quad s=0,1,...,L\ , \label{struct}%
\end{equation}
where the ghost $C^{\alpha_{s}|a_{0}...a_{s}}$ and auxiliary $B^{\alpha
_{s}|a_{1}...a_{s}}$ fields form symmetric $\mathrm{Sp}\left(  2\right)  $
tensors, being irreducible representations of the $\mathrm{Sp}\left(
2\right)  $ group, with the corresponding distribution \cite{BLT2} of the
Grassmann parity and ghost number. These fields absorb the pyramids of
ghost-antighost and Nakanishi--Lautrup fields of a given (ir)reducible gauge
theory, where $L$ in (\ref{struct}) is the corresponding stage of reducibility
\cite{BV}, and $L=0$ stands for an irreducible theory.

In the space of fields and antifields $(\phi^{A},\phi_{Aa}^{\ast},\bar{\phi
}_{A})$, one introduces the basic object of the BRST-antBRST Lagrangian
scheme, being an even-valued functional $S=S(\phi,\phi^{\ast},\bar{\phi})$
subject to an $\mathrm{Sp}\left(  2\right)  $-doublet of the generating
equations \cite{BLT1}%
\begin{equation}
\frac{1}{2}(S,S)^{a}+V^{a}S=i\hbar\Delta^{a}S\ \Longleftrightarrow\bar{\Delta
}^{a}\exp\left[  \left(  i/\hbar\right)  S\right]  =0\ ,\ \ \ \bar{\Delta}%
^{a}=\Delta^{a}+\left(  i/\hbar\right)  V^{a}\ . \label{3.3}%
\end{equation}
Here, $\hbar$ is the Planck constant, whereas the extended antibracket
$(\cdot,\cdot)^{a}$ and the operators $\Delta^{a}$, $V^{a}$ are given by%
\begin{equation}
(\cdot,\cdot)^{a}=\frac{\delta_{r}\cdot}{\delta\phi^{A}}\frac{\delta_{l}\cdot
}{\delta\phi_{Aa}^{\ast}}-\frac{\delta_{r}\cdot}{\delta\phi_{Aa}^{\ast}}%
\frac{\delta_{l}\cdot}{\delta\phi^{A}}\ ,\ \ \ \Delta^{a}=(-1)^{\varepsilon
_{A}}\frac{\delta_{l}}{\delta\phi^{A}}\frac{\delta}{\delta\phi_{Aa}^{\ast}%
}\ ,\ \ \ V^{a}=\varepsilon^{ab}\phi_{Ab}^{\ast}\frac{\delta}{\delta\bar{\phi
}_{A}}\ . \label{abrack}%
\end{equation}
The properties of the operators $\Delta^{a}$, $V^{a}$, $\bar{\Delta}^{a}$\ and
those of the extended antibracket $(\cdot,\cdot)^{a}$ were investigated in
\cite{BLT1}. The study of \cite{BLT3} proved the existence of solutions to
(\ref{3.3}) with the boundary condition $\left.  S\right\vert _{{\phi^{\ast
}=\bar{\phi}=\hbar=0}}=S_{0}\ $ in the form of an expansion in powers of
$\hbar$ and described the arbitrariness in solutions, which is controlled by a
transformation generated by the operators $\bar{\Delta}^{a}$, connecting two
solutions and describing the gauge-fixing procedure. A solution $S=S(\phi
,\phi^{\ast},\bar{\phi})$ of the generating equations (\ref{3.3}) allows one
to construct an extended (due to the antifields) generating functional of
Green's functions $Z\left(  J,\phi^{\ast},\bar{\phi}\right)  $ for the fields
$\phi^{A}$ of the total configuration space \cite{BLT1}, namely,%
\begin{equation}
Z\left(  J,\phi^{\ast},\bar{\phi}\right)  =\int d\phi\exp\left\{  \frac
{i}{\hbar}\left[  S_{\mathrm{ext}}\left(  \phi,\phi^{\ast},\bar{\phi}\right)
+J_{A}\phi^{A}\right]  \right\}  \ . \label{extZ}%
\end{equation}
Hence, the generating functional of Green's functions $Z(J)=\left.  Z\left(
J,\phi^{\ast},\bar{\phi}\right)  \right\vert _{{\phi^{\ast}=\bar{\phi}=0}}$ is
given by%
\begin{equation}
Z\left(  J\right)  =\int d\phi\exp\left\{  \frac{i}{\hbar}\left[
S_{\mathrm{eff}}\left(  \phi\right)  +J_{A}\phi^{A}\right]  \right\}
\,,\,\,\, \mathrm{with}\,\,\,S_{\mathrm{eff}}\left(  \phi\right)  =\left.
S_{\mathrm{ext}}\left(  \phi,\phi^{\ast},\bar{\phi}\right)  \right\vert
_{{\phi^{\ast}=\bar{\phi}=0}}\ , \label{genZ}%
\end{equation}
where $J_{A}$, $\varepsilon(J_{A})=\varepsilon_{A}$, are external sources to
the fields $\phi^{A}$, and $S_{\mathrm{ext}}=S_{\mathrm{ext}}\left(  \phi
,\phi^{\ast},\bar{\phi}\right)  $ is an action constructed with the help of an
even-valued gauge-fixing functional $F=F(\phi)$:%
\begin{equation}
\exp\left[  \left(  i/\hbar\right)  S_{\mathrm{ext}}\right]  =\hat{U}%
\exp\left[  \left(  i/\hbar\right)  S\right]  \,,\,\,\,\mathrm{with}%
\,\,\,\hat{U}=\exp\left(  F_{,A}\frac{\delta}{\delta\bar{\phi}_{A}}%
+\frac{i\hbar}{2}\varepsilon_{ab}\frac{\delta}{\delta\phi_{Aa}^{\ast}}%
F_{,AB}\frac{\delta}{\delta\phi_{Bb}^{\ast}}\right)  . \label{U}%
\end{equation}
Due to the commutativity of $\bar{\Delta}^{a}$ and $\hat{U}$, the gauge-fixing
procedure retains the form of the generating equations (\ref{3.3}),%
\begin{equation}
\bar{\Delta}^{a}\exp\left[  \left(  i/\hbar\right)  S_{\mathrm{ext}}\right]
=0\ . \label{ext}%
\end{equation}
A possible choice of the gauge-fixing functional $F(\phi)$\ has the form of
the most general $\mathrm{Sp}\left(  2\right)  $-scalar being quadratic in the
ghost and auxiliary fields \cite{BLT2}.

Introducing a set of auxiliary fields $\pi^{Aa}$ and $\lambda^{A}$,%
\begin{equation}
\varepsilon(\pi^{Aa})=\varepsilon_{A}+1\ ,\quad\varepsilon(\lambda
^{A})=\varepsilon_{A}\ ,\quad\mathrm{gh}(\pi^{Aa})=-(-1)^{a}+\mathrm{gh}%
(\phi^{A}),\quad\mathrm{gh}(\lambda^{A})=\mathrm{gh}(\phi^{A})\ ,
\label{pilambda}%
\end{equation}
one can represent $Z(J)$ as a functional integral in the extended space of
variables \cite{BLT1}%
\begin{equation}
Z(J)=\int d\Gamma\;\exp\left\{  \frac{i}{\hbar}\left[  S+\phi_{Aa}^{\ast}%
\pi^{Aa}+\left(  \bar{\phi}_{A}-F_{,A}\right)  \lambda^{A}-\left(  1/2\right)
\varepsilon_{ab}\pi^{Aa}F_{,AB}\pi^{Bb}+J_{A}\phi^{A}\right]  \right\}  \,,
\label{3.6}%
\end{equation}
where $d\Gamma=d\phi\ d\phi^{\ast}\ d\bar{\phi}\ d\lambda\ d\pi$ is the
integration measure.

An important property of the integrand in (\ref{3.6}) for $J_{A}=0$ is its
invariance under the following infinitesimal transformations of global
supersymmetry:%
\begin{equation}
\delta\left(  \phi^{A},\ \phi_{Aa}^{\ast},\ \bar{\phi}_{A},\ \pi
^{Aa},\ \lambda^{A}\right)  =\left(  \pi^{Aa}\mu_{a}\ ,\ \mu_{a}%
S_{,A}\ ,\ \varepsilon^{ab}\mu_{a}\phi_{Ab}^{\ast}\ ,\ -\varepsilon
^{ab}\lambda^{A}\mu_{b}\ ,\ 0\right)  \ , \label{3.7}%
\end{equation}
where $\mu_{a}$ is a doublet of constant anticommuting Grassmann parameters,
$\mu_{a}\mu_{b}+\mu_{b}\mu_{a}\equiv0$. The transformations (\ref{3.7})
realize the BRST-antiBRST transformations in the extended space ($\phi^{A}%
$,$\ \phi_{Aa}^{\ast}$,$\,\ \bar{\phi}_{A}$,$\ \pi^{Aa}$,$\ \lambda^{A}$).

The symmetry of the integrand in (\ref{3.6}) for $J_{A}=0$ under the
transformations (\ref{3.7}) with constant infinitesimal $\mu_{a}$ allows one
to derive the following Ward identities in the extended space:%
\begin{align}
&  J_{A}\langle\pi^{Aa}\rangle_{F,J} =0\,,\label{WI}\\
&  \mathrm{for}\,\,\, \langle\mathcal{O}\rangle_{F,J} = Z^{-1}(J)\int
d\Gamma\;\mathcal{O} \exp\left\{  \frac{i}{\hbar}\left[  S+\phi_{Aa}^{\ast}%
\pi^{Aa}+\left(  \bar{\phi}_{A}-F_{,A}\right)  \lambda^{A}-\left(  1/2\right)
\varepsilon_{ab}\pi^{Aa}F_{,AB}\pi^{Bb}+J_{A}\phi^{A}\right]  \right\}
,\nonumber
\end{align}
where the expectation value of a functional $\mathcal{O}(\Gamma)$ is given in
the extended space parameterized by $\Gamma$
with a gauge $F(\phi)$ in the presence of external sources $J_{A}$. To obtain
(\ref{WI}), we subject (\ref{3.6}) to a change of variables $\Gamma
\rightarrow\Gamma+\delta\Gamma$ with $\delta\Gamma$ given by (\ref{3.7}) and
use the equations (\ref{3.3}) for $S$. At the same time, with allowance for
the equivalence theorem \cite{equiv}, the transformations (\ref{3.7}) permit
one to establish the independence of the $S$-matrix from the choice of a
gauge. Indeed, suppose $Z_{F}\equiv Z(0)$ and change the gauge, $F\rightarrow
F+\Delta F$, by an infinitesimal value $\Delta F$. In the functional integral
for $Z_{F+\Delta F}$ we now make the change of variables (\ref{3.7}). Then,
choosing the parameters $\mu_{a}$ as%
\begin{equation}
\mu_{a}=-\frac{i}{2\hbar}\varepsilon_{ab}\left(  \Delta F\right)  _{,A}%
\pi^{Ab}\ , \label{inffdpar}%
\end{equation}
we find that $Z_{F+\Delta F}=Z_{F}$, and therefore the $S$-matrix is gauge-independent.

For the purpose of a subsequent treatment of Yang--Mills theories, we need the
particular case of solutions to the generating equations (\ref{3.3}) given by
a functional $S=S\left(  \phi,\phi^{\ast},\bar{\phi}\right)  $ linear in the
antifields. Namely, we assume%
\begin{equation}
S=S_{0}+\phi_{Aa}^{\ast}X^{Aa}+\bar{\phi}_{A}Y^{A}\ , \label{linS}%
\end{equation}
which implies%
\begin{equation}
S_{0,i}X^{ia}=0\ ,\quad X_{,B}^{Aa}X^{Bb}=\varepsilon^{ab}Y^{A}\ ,\quad
Y_{,A}^{B}X^{Aa}=0\ ,\quad X_{,A}^{Aa}=0 \label{xy2}%
\end{equation}
and allows one to present $S$ in the form%
\begin{equation}
S=S_{0}+\phi_{Aa}^{\ast}\left(  s^{a}\phi^{A}\right)  -\frac{1}{2}\bar{\phi
}_{A}\left(  s^{2}\phi^{A}\right)  \ ,\quad s^{2}\equiv s_{a}s^{a}\ ,
\label{SYMsa}%
\end{equation}
where $s^{a}$ are generators of BRST-antiBRST transformations,%
\begin{equation}
\delta\phi^{A}=\left(  s^{a}\phi^{A}\right)  \mu_{a}\ ,\quad s^{a}\phi
^{A}=X^{Aa}\ , \label{brst-antibrst}%
\end{equation}
and $s^{2}$ are generators of mixed BRST-antiBRST transformations,%
\begin{equation}
\delta^{2}\phi^{A}=s^{a}\left(  s^{b}\phi^{A}\mu_{b}\right)  \mu_{a}=-\frac
{1}{2}\left(  s^{2}\phi^{A}\right)  \mu^{2}\ ,\quad s^{2}\phi^{A}%
=\varepsilon_{ab}X_{,B}^{Aa}X^{Bb}=-2Y^{A}\ . \label{mixed}%
\end{equation}
The explicit form of $X^{Aa}$ and $Y^{A}$ for theories of Yang--Mills type was
found in \cite{BLT1} and is given in Appendix~\ref{AppB}.

For a solution of (\ref{3.3}) linear in the antifields, integration in
(\ref{3.6}) over $\phi_{Aa}^{\ast}$, $\bar{\phi}_{A}$, $\pi^{Aa}$,
$\lambda^{A}$ is trivial \cite{BLT1}:%
\begin{equation}
Z(J)=\int d\phi\ \exp\left\{  \frac{i}{\hbar}\left[  S_{F}\left(  \phi\right)
+J_{A}\phi^{A}\right]  \right\}  \;. \label{z(j)}%
\end{equation}
where%
\begin{equation}
S_{F}\left(  \phi\right)  =S_{0}\left(  A\right)  +F_{,A}Y^{A}-\left(
1/2\right)  \varepsilon_{ab}X^{Aa}F_{,AB}X^{Bb}\ , \label{action}%
\end{equation}
which can also be established directly by inserting the solution (\ref{linS})
into (\ref{U}).

The quantum action $S_{F}\left(  \phi\right)  $ can be presented in terms of a
mixed BRST-antiBRST variation,%
\begin{equation}
S_{F}\left(  \phi\right)  =S_{0}\left(  A\right)  -(1/2)s^{2}F\left(
\phi\right)  ~, \label{variat}%
\end{equation}
where the operators $s^{a}$, acting on an arbitrary functional $V=V\left(
\phi\right)  $ of any Grassmann parity, define a BRST-antiBRST analogue of the
Slavnov variation, $s^{a}V=V_{,A}\left(  s^{a}\phi^{A}\right)  $. Thus defined
operators $s^{a}$ are anticommuting, $s^{a}s^{b}+s^{b}s^{a}\equiv0$, for any
$a,b=1,2$\textbf{,}%
\begin{equation}
s^{a}s^{b}V=\varepsilon^{ab}W\ ,\quad W\equiv(1/2)\varepsilon_{ab}%
V_{,BA}X^{Aa}X^{Bb}\left(  -1\right)  ^{\varepsilon_{B}}-V_{,A}Y^{A}\ ,\quad
s^{a}s^{b}V=\left(  1/2\right)  \varepsilon^{ab}s^{2}V\ ,\ \ \ W=\left(
1/2\right)  s^{2}V\ , \label{V}%
\end{equation}
and therefore nilpotent, $s^{a}s^{b}s^{c}\equiv0$, which proves the invariance
of $S_{F}\,$given by (\ref{variat}) under the infinitesimal transformations
(\ref{brst-antibrst}),%
\[
\delta S_{F}=\left(  S_{F}\right)  _{,A}\delta\phi^{A}=\left(  s^{a}%
S_{F}\right)  \mu_{a}=\left(  s^{a}S_{0}\right)  \mu_{a}-\frac{1}{2}\left(
s^{a}s^{2}F\right)  \mu_{a}=0\,,
\]
by virtue of the condition $s^{a}S_{0}=S_{0,i}X^{ia}=0$ from (\ref{xy2}),
being a consequence of the Noether identities (\ref{Nid}).

In view of the condition $X_{,A}^{Aa}=0$ from (\ref{xy2}), the integration
measure in (\ref{z(j)}) is also invariant under the transformations
(\ref{brst-antibrst}), which ensures the invariance of the integrand in
(\ref{z(j)}) for $J_{A}=0$ under (\ref{brst-antibrst}). By analogy with the
previous consideration, this allows one to establish the Ward identities for
$Z(J)$ in (\ref{z(j)}),%
\begin{equation}
J_{A}\langle s^{a}\phi^{A}\rangle_{F,J}=J_{A}\langle X^{Aa}(\phi)\rangle
_{F,J}=0\ \ \ \mathrm{for}\mathtt{\ \ \ }\langle\mathcal{O}\rangle
_{F,J}=Z^{-1}(J)\int d\phi\;\mathcal{O}(\phi)\exp\left\{  \frac{i}{\hbar
}\left[  S_{F}\left(  \phi\right)  +J_{A}\phi^{A}\right]  \right\}  \, ,
\label{WIl}%
\end{equation}
as well as the independence of the $S$-matrix from the choice of a gauge.
Indeed, suppose $Z_{F}\equiv Z(0)$ in (\ref{z(j)}) and change the gauge
$F\rightarrow F+\Delta F$ by an infinitesimal value $\Delta F$. Then, making
in $Z_{F+\Delta F}$ the change of variables (\ref{brst-antibrst}) with the
field-dependent infinitesimal parameters
\begin{equation}
\mu_{a}=\frac{i}{2\hbar}\varepsilon_{ab}\left(  \Delta F\right)  _{,A}%
X^{Ab}=\frac{i}{2\hbar}\left(  s_{a}\Delta F\right)  \,, \label{partic_case}%
\end{equation}
being a particular case of the field-dependent BRST-antiBRST\ transformations
studied in the following section, we find $Z_{F+\Delta F}=Z_{F}$, which
establishes the gauge-independence of the $S$-matrix.

\section{Finite BRST-antiBRST Transformations and their Jacobians}

\label{fBRSTa}\renewcommand{\theequation}{\arabic{section}.\arabic{equation}} \setcounter{equation}{0}

Let us introduce finite transformations of the fields $\phi^{A}$ with a
doublet $\lambda_{a}$ of anticommuting Grassmann parameters, $\lambda
_{a}\lambda_{b}+\lambda_{b}\lambda_{a}=0$,%
\begin{equation}
\phi^{A}\rightarrow\phi^{\prime A}=\phi^{A}+\Delta\phi^{A}=\phi^{\prime
A}\left(  \phi|\lambda\right)  \ ,\ \ \ \mathrm{so\ \ that}\ \ \ \phi^{\prime
A}\left(  \phi|0\right)  =\phi^{A}\ . \label{defin}%
\end{equation}
In the general case, such transformations are quadratic in the parameters, due
to $\lambda_{a}\lambda_{b}\lambda_{c}\equiv0$,%
\begin{equation}
\phi^{\prime A}\left(  \phi|\lambda\right)  =\phi^{\prime A}\left(
\phi|0\right)  +\left[  \frac{\overleftarrow{\partial}}{\partial\lambda_{a}%
}\phi^{\prime A}\left(  \phi|\lambda\right)  \right]  _{\lambda=0}\lambda
_{a}+\frac{1}{2}\left[  \frac{\overleftarrow{\partial}}{\partial\lambda_{a}%
}\frac{\overleftarrow{\partial}}{\partial\lambda_{b}}\phi^{\prime A}\left(
\phi|\lambda\right)  \right]  \lambda_{a}\lambda_{b}\ , \label{Z}%
\end{equation}
which implies%
\begin{equation}
\Delta\phi^{A}=Z^{Aa}\lambda_{a}+\left(  1/2\right)  Z^{A}\lambda
^{2}\,,\,\,\,\mathrm{where}\,\,\,\, \lambda^{2}\equiv\lambda_{a}\lambda^{a}\ ,
\label{Z2}%
\end{equation}
for certain functions $Z^{Aa}=Z^{Aa}\left(  \phi\right)  $, $Z^{A}%
=Z^{A}\left(  \phi\right)  $, corresponding to the first- and second-order
derivatives of $\phi^{\prime A}\left(  \phi|\lambda\right)  $ with respect to
$\lambda_{a}$ in (\ref{Z}).

In view of the obvious property of nilpotency $\Delta\phi^{A_{1}}\cdots
\Delta\phi^{A_{n}}\equiv0$, $n\geq3$, an arbitrary functional $F\left(
\phi\right)  $ under the above transformations $\phi^{A}\rightarrow\phi
^{A}+\Delta\phi^{A}$ can be expanded as%
\begin{equation}
F\left(  \phi+\Delta\phi\right)  =F\left(  \phi\right)  +F_{,A}\left(
\phi\right)  \Delta\phi^{A}+\left(  1/2\right)  F_{,AB}\left(  \phi\right)
\Delta\phi^{B}\Delta\phi^{A}\ . \label{F}%
\end{equation}

Based on (\ref{defin})--(\ref{F}), we now introduce \emph{finite BRST-antiBRST
transformations} as invariance transformations of the quantum action
$S_{F}\left(  \phi\right)  $ given by (\ref{variat}) under finite
transformations of the fields $\phi^{A}$, such that%
\begin{equation}
S_{F}\left(  \phi+\Delta\phi\right)  =S_{F}\left(  \phi\right)
\ ,\ \ \ \left[  \frac{\overleftarrow{\partial}}{\partial\lambda_{a}}%
\Delta\phi^{A}\right]  _{\lambda=0}=s^{a}\phi^{A}\ \ \mathrm{and}%
\mathtt{\ \ }\left[  \frac{\overleftarrow{\partial}}{\partial\lambda_{a}}%
\frac{\overleftarrow{\partial}}{\partial\lambda_{b}}\Delta\phi^{A}\right]
=\frac{1}{2}\varepsilon^{ab}s^{2}\phi^{A}, \label{finBRSTantiBRST}%
\end{equation}
which implies $Z^{Aa}=s^{a}\phi^{A}=X^{Aa}$ and $Z^{A}=\left(  1/2\right)
s^{2}\phi^{A}=-Y^{A}$, according to (\ref{brst-antibrst}), (\ref{mixed}),
(\ref{Z2}).

One can easily verify the consistency of definition (\ref{finBRSTantiBRST}) by
considering the equation, implied by $\Delta S_{F}=0$,%
\begin{equation}
\left(  S_{F}\right)  _{,A}\left(  X^{Aa}\lambda_{a}-\frac{1}{2}Y^{A}%
\lambda^{2}\right)  +\frac{1}{2}\left(  S_{F}\right)  _{,AB}\left(
X^{Bb}\lambda_{b}-\frac{1}{2}Y^{B}\lambda^{2}\right)  \left(  X^{Aa}%
\lambda_{a}-\frac{1}{2}Y^{A}\lambda^{2}\right)  =0\ . \label{compbab}%
\end{equation}
Taking into account the fact $\lambda_{a}\lambda^{2}=\lambda^{4}\equiv0$, the
invariance relations $\left(  S_{F}\right)  _{,A}X^{Aa}=0$, and their
differential consequences $\left(  S_{F}\right)  _{,AB}X^{Bb}\lambda_{b}%
X^{Aa}\lambda_{a}=\left(  S_{F}\right)  _{,A}Y^{A}\lambda^{2}$, implied by the
relations $Y^{A}=\left(  1/2\right)  X_{,B}^{Aa}X^{Bb}\varepsilon_{ba}$ from
(\ref{mixed}), we find that the above equation is satisfied identically:%
\begin{equation}
\left(  S_{F}\right)  _{,A}X^{Aa}\lambda_{a}-\frac{1}{2}\left(  S_{F}\right)
_{,A}Y^{A}\lambda^{2}+\frac{1}{2}\left(  S_{F}\right)  _{,AB}X^{Bb}\lambda
_{b}X^{Aa}\lambda_{a}\equiv0\ .\nonumber
\end{equation}
Explicitly, the finite BRST-antiBRST transformations can be
presented as \footnote{Finite BRST-antiBRST transformations
(\ref{finite}) may be regarded as an extension of finite ``mixed
BRST'' transformations \cite{Upadhyay3}, which include only the
linear dependence on odd-valued parameters $\Theta_1$, $\Theta_2$;
see Introduction for details.}
\begin{equation}
\Delta\phi^{A}=X^{Aa}\lambda_{a}-\frac{1}{2}Y^{A}\lambda^{2}=\left(  s^{a}%
\phi^{A}\right)  \lambda_{a}+\frac{1}{4}\left(  s^{2}\phi^{A}\right)
\lambda^{2}\ , \label{finite}%
\end{equation}
which implies that the finite variation $\Delta\phi^{A}$ includes
the generators of BRST-antiBRST transformations $\left(
s^{1},s^{2}\right)  $, as well as their commutator
$s^{2}=\varepsilon_{ab}s^{b}s^{a}=s^{1}s^{2}%
-s^{2}s^{1}$.

According to (\ref{V}), (\ref{F}), (\ref{finite}) and $\lambda_{a}\lambda
^{2}=\lambda^{4}\equiv0$, the variation $\Delta F\left(  \phi\right)  $ of an
arbitrary functional $F\left(  \phi\right)  $ under the finite BRST-antiBRST
transformations is given by%
\begin{align}
\Delta F  &  =F_{,A}X^{Aa}\lambda_{a}-\frac{1}{2}F_{,A}Y^{A}\lambda^{2}%
+\frac{1}{2}F_{,AB}X^{Bb}\lambda_{b}X^{Aa}\lambda_{a}\nonumber\\
&  =\left(  F_{,A}X^{Aa}\right)  \lambda_{a}+\frac{1}{2}\left(  \frac{1}%
{2}\varepsilon_{ab}F_{,BA}X^{Aa}X^{Bb}\left(  -1\right)  ^{\varepsilon_{B}%
}-F_{,A}Y^{A}\right)  \lambda^{2}=\left(  s^{a}F\right)  \lambda_{a}+\frac
{1}{4}\left(  s^{2}F\right)  \lambda^{2}\ . \label{DeltaF}%
\end{align}
This relation allows one to study the group properties of finite BRST-antiBRST
transformations (\ref{finite}), with account taken for the fact that these
transformations do not form a Lie superalgebra, nor a vector superspace
structure, due to the presence of the term which is quadratic in $\lambda_{a}%
$. Namely, we have (for details, see Appendix~\ref{AppC})%
\begin{equation}
\Delta_{\left(  1\right)  }\Delta_{\left(  2\right)  }F=\left(  s^{a}%
\Delta_{\left(  2\right)  }F\right)  \lambda_{\left(  1\right)  a}+\frac{1}%
{4}\left(  s^{2}\Delta_{\left(  2\right)  }F\right)  \lambda_{\left(
1\right)  }^{2}\equiv\left(  s^{a}F\right)  \vartheta_{\left(  1,2\right)
a}+\frac{1}{4}\left(  s^{2}F\right)  \theta_{\left(  1,2\right)  }\ ,
\label{D1D2F}%
\end{equation}
for certain functionals $\vartheta_{\left(  1,2\right)  }^{a}=\vartheta
_{\left(  1,2\right)  }^{a}\left(  \phi\right)  $ and $\theta_{\left(
1,2\right)  }=\theta_{\left(  1,2\right)  }\left(  \phi\right)  $, constructed
explicitly in (\ref{vartheta12}), (\ref{theta12}) from the parameters of
finite transformations, which are generally field-dependent, $\lambda_{\left(
j\right)  }^{a}=\lambda_{\left(  j\right)  }^{a}\left(  \phi\right)  $, for
$j=1,2 $. Therefore, the commutator of finite variations has the form%
\begin{equation}
\left[  \Delta_{\left(  1\right)  },\Delta_{\left(  2\right)  }\right]
F=\left(  s^{a}F\right)  \vartheta_{\left[  1,2\right]  a}+\frac{1}{4}\left(
s^{2}F\right)  \theta_{\left[  1,2\right]  }\ ,\ \ \ \vartheta_{\left[
1,2\right]  }^{a}\equiv\vartheta_{\left(  1,2\right)  }^{a}-\vartheta_{\left(
2,1\right)  }^{a}\ \ ,\ \ \ \theta_{\left[  1,2\right]  }\equiv\theta_{\left(
1,2\right)  }-\theta_{\left(  2,1\right)  }\ , \label{[D1D2]}%
\end{equation}
where $\vartheta_{\left[  1,2\right]  }^{a}$\textbf{, }$\theta_{\left[
1,2\right]  }$ are given explicitly by (\ref{vartheta[12]}), (\ref{theta[12]})
and possess the symmetry properties $\vartheta_{\left[  1,2\right]  }%
^{a}=-\vartheta_{\left[  2,1\right]  }^{a}$, $\theta_{\left[  1,2\right]
}=-\theta_{\left[  2,1\right]  }$. In particular, assuming $F\left(
\phi\right)  =\phi^{A}$ in (\ref{[D1D2]}), we have
\begin{equation}
\left[  \Delta_{\left(  1\right)  },\Delta_{\left(  2\right)  }\right]
\phi^{A}=\left(  s^{a}\phi^{A}\right)  \vartheta_{\left[  1,2\right]  a}%
+\frac{1}{4}\left(  s^{2}\phi^{A}\right)  \theta_{\left[  1,2\right]  }\ .
\label{[D1D2]Ph}%
\end{equation}
In general, the commutator (\ref{[D1D2]Ph}) of finite non-linear
transformations (\ref{finite}) does not belong to the class of these
transformations, due to the opposite symmetry properties of $\vartheta
_{\left[  1,2\right]  a}\vartheta_{\left[  1,2\right]  }^{a}$ and
$\theta_{\left[  1,2\right]  }$,%
\begin{equation}
\vartheta_{\left[  1,2\right]  a}\vartheta_{\left[  1,2\right]  }%
^{a}=\vartheta_{\left[  2,1\right]  a}\vartheta_{\left[  2,1\right]  }%
^{a}\ ,\ \ \ \theta_{\left[  1,2\right]  }=-\theta_{\left[  2,1\right]  }\ ,
\label{propfbab}%
\end{equation}
which reflects the fact that a finite BRST-antiBRST transformation looks as a
group element, i.e., not as an element of a Lie superalgebra; however, the
linear approximation $\Delta^{\mathrm{lin}}\phi^{A}=\left(  s^{a}\phi
^{A}\right)  \lambda_{a}$ to a finite transformation $\Delta\phi^{A}%
=\Delta^{\mathrm{lin}}\phi^{A}+O\left(  \lambda^{2}\right)  $ does form an
algebra. Indeed, due to (\ref{[d1d2]F}), (\ref{vartheta[12]}),
(\ref{theta[12]}), we have%
\begin{equation}
\left[  \Delta_{\left(  1\right)  }^{\mathrm{lin}},\Delta_{\left(  2\right)
}^{\mathrm{lin}}\right]  F=\Delta_{\left[  1,2\right]  }^{\mathrm{lin}%
}F=\left(  s^{a}F\right)  \lambda_{\left[  1,2\right]  a}\ ,\ \ \ \lambda
_{\left[  1,2\right]  }^{a}\equiv\left(  s_{b}\lambda_{\left(  1\right)  }%
^{a}\right)  \lambda^{b}_{\left(  2\right)  }-\left(  s_{b}\lambda_{\left(
2\right)  }^{a}\right)  \lambda^{b}_{\left(  1\right)  }\ . \label{comm}%
\end{equation}
Thus, the construction of finite BRST-antiBRST transformations (\ref{finite})
reduces to the usual BRST-antiBRST transformations (\ref{brst-antibrst}),
$\delta\phi^{A}=\Delta^{\mathrm{lin}}\phi^{A}$, linear in the infinitesimal
parameter $\mu_{a}=\lambda_{a}$, as one selects in (\ref{finite}) the
approximation that forms an algebra with respect to the commutator.

Let us now consider the modification of the integration measure $d\phi
\rightarrow d\phi^{\prime}$ in (\ref{z(j)}) under the finite transformations
$\phi^{A}\rightarrow\phi^{\prime A}=\phi^{A}+\Delta\phi^{A}$, with $\Delta
\phi^{A}$ given by (\ref{finite}),%
\begin{equation}
d\phi^{\prime}=d\phi\ \mathrm{Sdet}\left(  \frac{\delta\phi^{\prime}}%
{\delta\phi}\right)  ,\,\,\,\mathrm{\ with }\,\,\, \mathrm{Sdet}\left(
\frac{\delta\phi^{\prime}}{\delta\phi}\right)  =\mathrm{Sdet}\left(
\mathbb{I}+M\right)  =\exp\left[  \mathrm{Str}\ln\left(  \mathbb{I}+M\right)
\right]  \equiv\exp\left(  \Im\right)  \ , \label{measure}%
\end{equation}
where the Jacobian $\exp\left(  \Im\right)  $ has the form%
\begin{equation}
\Im=\mathrm{Str}\ln\left(  \mathbb{I}+M\right)  =-\sum_{n=1}^{\infty}%
\frac{\left(  -1\right)  ^{n}}{n}\,\,\mathrm{Str}\left(  M^{n}\right)
,\,\,\,\mathrm{\ for }\,\,\, \mathrm{Str}\left(  M^{n}\right)  =\left(
M^{n}\right)  _{A}^{A}\left(  -1\right)  ^{\varepsilon_{A}}\,\,\,\mathrm{\ and
}\,\,\,M_{B}^{A}\equiv\frac{\delta\left(  \Delta\phi^{A}\right)  }{\delta
\phi^{B}}\ . \label{superJ}%
\end{equation}

In the case of \emph{global} finite transformations, corresponding to
$\lambda_{a}=\mathrm{const}$, the integration measure remains invariant (for
details, see Appendix~\ref{AppA1})
\begin{equation}
\Im\left(  \phi\right)  =0\ \Longrightarrow\ \left(  \mathrm{Sdet}\left(
\frac{\delta\phi^{\prime}}{\delta\phi}\right)  =1\,\,\,\mathrm{and}%
\,\,\,d\phi^{\prime}=d\phi\right)  \ . \label{constJ}%
\end{equation}
Due to the invariance of the quantum action $S_{F}=S_{0}+\left(  1/2\right)
s^{a}s_{a}F$ under $\phi^{A}\rightarrow\phi^{\prime A}$ the above implies that
the integrand with the vanishing sources $\mathcal{I}_{\phi}\equiv d\phi
\exp\left[  \left(  i/\hbar\right)  S_{F}\right]  $ in (\ref{z(j)}) is also
invariant, $\mathcal{I}_{\phi^{\prime}}=\mathcal{I}_{\phi}$, under the
transformations (\ref{finite}), which justifies their interpretation as finite
BRST-antiBRST transformations.

As we turn to finite \emph{field-dependent} transformations, let us examine
the particular case\footnote{\label{4}Notice that the parameters $\lambda_{a}%
$, $a=1,2$, in the case $\lambda_{a}=s_{a}\Lambda$ are not functionally
independent: $s^{1}\lambda_{1}+s^{2}\lambda_{2}=-s^{2}\Lambda$.} $\lambda
_{a}\left(  \phi\right)  =s_{a}\Lambda\left(  \phi\right)  $ with a certain
even-valued potential, $\Lambda=\Lambda\left(  \phi\right)  $, which is
inspired by infinitesimal field-dependent BRST-antiBRST transformations with
the parameters (\ref{partic_case}). In this case, the integration measure
takes the form (relation (\ref{superJ1}) is deduced in Appendix~\ref{AppA2})%
\begin{align}
&  \Im\left(  \phi\right)  \ =\ -2\mathrm{\ln}\left[  1+f\left(  \phi\right)
\right]  ,\,\,\,\mathrm{\ with }\,\,\, f\left(  \phi\right)  =-\frac{1}%
{2}s^{2} \Lambda\left(  \phi\right)  ,\,\,\,\mathrm{\ for }\,\,\,s^{a}%
s_{a}=-s^{2}\ ,\label{superJaux}\\
&  d\phi^{\prime}\ =\ d\phi\ \exp\left[  \frac{i}{\hbar} \left(  -i\hbar
\Im\right)  \right]  =d\phi\ \exp\left\{  \frac{i}{\hbar}\left[
i\hbar\,\mathrm{\ln}\left(  1+\frac{1}{2}s^{a}s_{a}\Lambda\right)
^{2}\right]  \right\}  \ . \label{superJ1}%
\end{align}
In view of the invariance of the quantum action $S_{F}\left(  \phi\right)  $
under (\ref{finite}), the change $\phi^{A}\rightarrow\phi^{\prime A}=\phi
^{A}+\Delta\phi^{A}$ induces in (\ref{z(j)}) the following transformation of
the integrand with the vanishing sources, $\mathcal{I}_{\phi}\equiv d\phi
\exp\left[  \left(  i/\hbar\right)  S_{F}\left(  \phi\right)  \right]  $:%
\begin{equation}
\mathcal{I}_{\phi+\Delta\phi}\ =\ d\phi\ \exp\left[  \Im\left(  \phi\right)
\right]  \exp\left[  \left(  i/\hbar\right)  S_{F}\left(  \phi+\Delta
\phi\right)  \right]  \ =\ d\phi\ \exp\left\{  \left(  i/\hbar\right)  \left[
S_{F}\left(  \phi\right)  -i\hbar\Im\left(  \phi\right)  \right]  \right\}
\ , \label{superJ2}%
\end{equation}
whence%
\begin{equation}
\mathcal{I}_{\phi+\Delta\phi}\ =\ d\phi\ \exp\left\{  \left(  i/\hbar\right)
\left[  S_{F}\left(  \phi\right)  +i\hbar\ \mathrm{\ln}\left(  1+s^{a}%
s_{a}\Lambda\left(  \phi\right)  /2\right)  ^{2}\right]  \right\}  \ .
\label{superJ3}%
\end{equation}
Due to the explicit form of the initial quantum action $S_{F}=S_{0}+\left(
1/2\right)  s^{a}s_{a}F$, the BRST-antiBRST-exact contribution $i\hbar\,
\mathrm{\ln}\left(  1+s^{a}s_{a}\Lambda/2\right)  ^{2}$ to the action $S_{F}$,
resulting from the transformation of the integration measure, can be
interpreted as a change of the gauge-fixing functional made in the original
integrand $\mathcal{I}_{\phi}$,%
\begin{align}
&  i\hbar\ \mathrm{\ln}\left(  1+s^{a}s_{a}\Lambda/2\right)  ^{2}%
\ =\ s^{a}s_{a}\left(  \Delta F/2\right) \label{superJ3m}\\
&  \Longrightarrow\ \ \ \mathcal{I}_{\phi+\Delta\phi}\ =\ d\phi\ \exp\left\{
\left(  i/\hbar\right)  \left[  S_{0}+\left(  1/2\right)  s^{a}s_{a}\left(
F+\Delta F\right)  \right]  \right\}  =\left.  \mathcal{I}_{\phi}\right\vert
_{F\rightarrow F+\Delta F}\ , \label{superJ4}%
\end{align}
for a certain $\Delta F\left(  \phi\right)  $, whose relation to
$\Lambda\left(  \phi\right)  $ is discussed below. In other words, the
field-dependent transformations with the parameters $\lambda_{a}=s_{a}\Lambda$
amount to a\emph{\ precise change of the gauge-fixing functional}. As a
consequence, the integrand in (\ref{z(j)}) for $J_{A}=0$, corresponding to the
quantum action $S_{F+\Delta F}=S_{0}+\left(  1/2\right)  s^{a}s_{a}\left(
F+\Delta F\right)  $ with a modified gauge-fixing functional, is invariant
under both the infinitesimal, $\delta\phi^{A}$, and finite, $\Delta\phi^{A}$,
BRST-antiBRST transformations, with constant parameters $\mu_{a}$ and
$\lambda_{a}$ in (\ref{brst-antibrst}) and (\ref{finite}), respectively.

Let us denote by $T^{\left(  \Delta F\right)  }$ the operation that transforms
an integrand $\mathcal{I}_{\phi}^{\left(  F\right)  }$ into $\mathcal{I}%
_{\phi}^{\left(  F+\Delta F\right)  }$, corresponding to the respective
gauge-fixing functionals $F$ and $F+\Delta F$,%
\begin{equation}
T^{\left(  \Delta F\right)  }:\ \ \mathcal{I}_{\phi}^{\left(  F\right)
}\mathcal{\rightarrow I}_{\phi}^{\left(  F+\Delta F\right)  }\ , \label{TF}%
\end{equation}
which implies an additive composition law:%
\begin{equation}
T^{\left(  \Delta F_{1}\right)  }\circ T^{\left(  \Delta F_{2}\right)
}=T^{\left(  \Delta F_{2}\right)  }\circ T^{\left(  \Delta F_{1}\right)
}=T^{\left(  \Delta F_{1}+\Delta F_{2}\right)  }\ . \label{TFcomp}%
\end{equation}
As we denote by $\Lambda^{\left(  \Delta F\right)  }$ the gauge-fixing
functional corresponding to $\Delta F$, there follow the properties%
\begin{equation}
\mathrm{\ln}\left(  1+s^{a}s_{a}\Lambda^{\left(  \Delta F_{1}+\Delta
F_{2}\right)  }/2\right)  ^{2}=\mathrm{\ln}\left(  1+s^{a}s_{a}\Lambda
^{\left(  \Delta F_{1}\right)  }/2\right)  ^{2}+\mathrm{\ln}\left(
1+s^{a}s_{a}\Lambda^{\left(  \Delta F_{2}\right)  }/2\right)  ^{2}%
\ ,\ \ \ \Lambda^{\left(  0\right)  }=0\ , \label{TLF}%
\end{equation}
implying relations between $s^{2}\Lambda^{\left(  \Delta F_{1}+\Delta
F_{2}\right)  }$ and $s^{2}\Lambda^{\left(  \Delta F_{j}\right)  }$ for $j=1$,
$2$, as well as between $s^{2}\Lambda^{\left(  -\Delta F\right)  }$ and
$s^{2}\Lambda^{\left(  \Delta F\right)  }$:%
\begin{align}
s^{2}\Lambda^{\left(  \Delta F_{1}+\Delta F_{2}\right)  }  &  =s^{2}\left(
\Lambda^{\left(  \Delta F_{1}\right)  }+\Lambda^{\left(  \Delta F_{2}\right)
}\right)  -\left(  s^{2}\Lambda^{\left(  \Delta F_{1}\right)  }\right)
\left(  s^{2}\Lambda^{\left(  \Delta F_{2}\right)  }\right)
/2\ ,\label{LF1F2}\\
s^{2}\Lambda^{\left(  -\Delta F\right)  }  &  =-\left(  s^{2}\Lambda^{\left(
\Delta F\right)  }\right)  \left[  1-\left(  s^{2}\Lambda^{\left(  \Delta
F\right)  }\right)  /2\right]  ^{-1}\ . \label{L-FF}%
\end{align}

The relation (\ref{superJ3m}) between the potential $\Lambda\left(
\phi\right)  $ and the variation $\Delta F\left(  \phi\right)  $ of the
gauge-fixing functional can be considered as a compensation equation (for the
unknown functional $\Delta F\left(  \phi\right)  $, with a given
$\Lambda\left(  \phi\right)  $, and vice versa),
\begin{equation}
i\hbar\ \ln\left(  1+s^{a}s_{a}\Lambda\left(  \phi\right)  /2\right)
^{2}=s^{a}s_{a}\Delta F\left(  \phi\right)  /2\ , \label{Lambda-F}%
\end{equation}
whose solution, up to BRST-antiBRST-exact terms, has the form%
\begin{equation}
\Delta F\left(  \phi\right)  =2i\hbar\ \Lambda\left(  \phi\right)  \left(
s^{a}s_{a}\Lambda\left(  \phi\right)  \right)  ^{-1}\ln\left(  1+s^{a}%
s_{a}\Lambda\left(  \phi\right)  /2\right)  ^{2}\ . \label{Lambda-Fsol}%
\end{equation}
The relation (\ref{Lambda-F}) can be inverted as an equation for
$\Lambda\left(  \phi\right)  $, namely,%
\begin{equation}
s^{a}s_{a}\Lambda=2\left[  \exp\left(  \frac{1}{4i\hbar}s^{a}s_{a}\Delta
F\right)  -1\right]  \,. \label{Lambda-F1}%
\end{equation}
Up to BRST-antiBRST-exact terms, its solution reads%
\begin{equation}
\Lambda=2\Delta F\left(  s^{a}s_{a}\Delta F\right)  ^{-1}\left[  \exp\left(
\frac{1}{4i\hbar}s^{b}s_{b}\Delta F\right)  -1\right]  =\frac{1}{2i\hbar
}\Delta F\sum_{n=0}^{\infty}\frac{1}{\left(  n+1\right)  !}\left(  \frac
{1}{4i\hbar}s^{a}s_{a}\Delta F\right)  ^{n}\ , \label{Lambda-Fsol1}%
\end{equation}
whence%
\begin{align}
\lambda_{a}  &  =s_{a}\Lambda=\frac{1}{2i\hbar}\left(  s_{a}\Delta F\right)
\sum_{n=0}^{\infty}\frac{1}{\left(  n+1\right)  !}\left(  \frac{1}{4i\hbar
}s^{b}s_{b}\Delta F\right)  ^{n}\nonumber\\
&  =\frac{1}{2i\hbar}\left(  s_{a}\Delta F\right)  \left[  1+\frac{1}%
{2!}\left(  \frac{1}{4i\hbar}s^{b}s_{b}\Delta F\right)  +\frac{1}{3!}\left(
\frac{1}{4i\hbar}s^{b}s_{b}\Delta F\right)  ^{2}+\frac{1}{4!}\left(  \frac
{1}{4i\hbar}s^{b}s_{b}\Delta F\right)  ^{3}+\ldots\right]  \,. \label{2}%
\end{align}
In particular, the first order of $\lambda_{a}=\mu_{a}$ in powers of $\Delta
F$ has the form%
\begin{equation}
\mu_{a}=-\frac{i}{2\hbar}\left(  s_{a}\Delta F\right)  \ . \label{linear}%
\end{equation}

Using (\ref{2}), one can construct a finite BRST-antiBRST transformation that
connects two quantum theories of Yang--Mills type corresponding to some
gauge-fixing functionals $F$ and $F+\Delta F$ for a given finite variation
$\Delta F$. The symmetry of the integrand in (\ref{z(j)}) for $J_{A}=0$ under
the transformations (\ref{finite}) allows one to establish the independence of
the $S$-matrix from the choice of a gauge. Indeed, suppose $Z_{F}\equiv Z(0)$
and change the gauge $F\rightarrow F+\Delta F$ by a finite value $\Delta F$.
In the functional integral for $Z_{F+\Delta F}$ we now make the change of
variables (\ref{finite}). Then, selecting the parameters $\lambda_{a}%
=s_{a}\Lambda$ to meet the condition%
\begin{equation}
i\hbar\ln\left(  1+s^{a}s_{a}\Lambda/2\right)  ^{2}=-\left(  1/2\right)
s^{a}s_{a}\Delta F\ , \label{meet}%
\end{equation}
cf. (\ref{Lambda-F}), we find that $Z_{F+\Delta F}=Z_{F}$, whence, due to the
equivalence theorem \cite{equiv}, the $S$-matrix is gauge-independent. In the
particular case of an infinitesimal variation $\Delta F$, condition
(\ref{meet}) produces, in virtue of (\ref{linear}), precisely the form
(\ref{partic_case}) of field-dependent parameters $\lambda_{a}=\mu_{a}$ in the
framework of infinitesimal BRST-antiBRST\ transformations.

As we identify $\lambda_{a}=s_{a}\Lambda$ with a solution of (\ref{Lambda-F}),
$\Lambda^{\left(  \Delta F\right)  }\equiv\Lambda\left(  \Delta F\right)  $,
the representation (\ref{z(j)}) describes the dependence of the functional
$Z_{F}(J)$ on a finite variation of the gauge:%
\begin{equation}
\Delta Z_{F}(J)=\frac{i}{\hbar}Z_{F}(J)\left\langle J_{A}\left[  (s^{a}%
\phi^{A})s_{a}\Lambda(-\Delta F)+\frac{1}{4}(s^{2}\phi^{A})\left[
s\Lambda(-\Delta F)\right]  ^{2}+\frac{i}{4\hbar}\varepsilon_{ab}(s^{a}%
\phi^{A})J_{B}(s^{b}\phi^{B})\left[  s\Lambda(-\Delta F)\right]  ^{2}\right]
\right\rangle _{F,J}\,, \label{gaugedepzj}%
\end{equation}
where $\Delta Z_{F}(J)\equiv Z_{F+\Delta F}(J)-Z_{F}(J)$. The above relation
(\ref{gaugedepzj}) generalizes the gauge-dependence of $Z(J)$ in Yang--Mills
type theories to the case of finite variations of the gauge.

\section{Correspondence between Gauges in Yang--Mills Theories}

\label{YMgauges}%
\renewcommand{\theequation}{\arabic{section}.\arabic{equation}} \setcounter{equation}{0}

In this section, we consider the Yang--Mills theory, given by the action%
\begin{equation}
S_{0}(A)=-\frac{1}{4}\int d^{D}x\ F_{\mu\nu}^{m}F^{m\mu\nu},
\,\,\,\,\mathrm{\ for }\,\,\,\, F_{\mu\nu}^{m}=\partial_{\mu}A_{\nu}%
^{m}-\partial_{\nu}A_{\mu}^{m}+f^{mnl}A_{\mu}^{n}A_{\nu}^{l}\,\,, \label{4.1}%
\end{equation}
with the Lorentz indices $\mu,\nu=0,1,\ldots,D{-}1$, the metric tensor
$\eta_{\mu\nu}=\mathrm{diag}(-,+,\ldots,+)$, and the totally
antisymmetric~$su(N)$ structure constants $f^{lmn}$ for $l,m,n = 1,\ldots,
N^{2}-1$.

The action (\ref{4.1}) is invariant under the gauge transformations
\begin{equation}
\delta A_{\mu}^{m}(x)=D_{\mu}^{mn}(x)\zeta^{n}(x)=\int d^{D}y\ R_{\mu}%
^{mn}(x;y)\zeta^{n}(y)\ ,\quad D_{\mu}^{mn}=\delta^{mn}\partial_{\mu}%
+f^{mln}A_{\mu}^{l}\,\,, \label{R(A)}%
\end{equation}
with arbitrary Bosonic functions $\zeta^{n}(y)$ in $\mathbb{R}^{1,D-1}$, the
covariant derivative $D_{\mu}^{mn}$, and the generators $R_{\mu}%
^{mn}(x;y)=R_{\alpha}^{i}$ of the gauge transformations, the condensed indices
being $i=(\mu,m,x)$, $\alpha=(n,y)$. The generators $R_{\alpha}^{i}$ in
(\ref{R(A)}) form a closed gauge algebra with $M_{\alpha\beta}^{ij}=0$ in
(\ref{gauge_alg}), whereas the structure coefficients $F_{\alpha\beta}%
^{\gamma}$ arising in (\ref{gauge_alg}) are given by%
\begin{equation}
F_{\alpha\beta}^{\gamma}=f^{lmn}\delta(x-z)\delta(y-z)\,,\,\,\, \mathrm{\ for
}\,\,\,\alpha=(m,x)\,, \ \beta=(n,y)\,,\ \gamma=(l,z)\,. \label{F(A)}%
\end{equation}

The total configuration space of fields $\phi^{A}$ and the corresponding
antifields $\phi_{Aa}^{\ast}$, $\bar{\phi}_{A}$ of the theory are given by%
\begin{equation}
\phi^{A}=\left(  A^{\mu m},B^{m},C^{ma}\right)  \ ,\quad\phi_{Aa}^{\ast
}=\left(  A_{\mu a}^{\ast m},B_{a}^{\ast m},C_{ab}^{\ast m}\right)
\,,\quad\bar{\phi}_{A}=\left(  \bar{A}_{\mu}^{m},\bar{B}^{m},\bar{C}_{a}%
^{m}\right)  \,. \label{YM-fields}%
\end{equation}
With allowance made for (\ref{antif}), the Grassmann parity and ghost number
assume the values
\begin{equation}
\varepsilon(\phi^{A})\equiv\left(  0,0,1\right)  \ ,\ \ \ \ \mathrm{gh}%
(\phi^{A})=\left(  0,0,\left(  -1\right)  ^{a+1}\right)  \,. \label{YM-e,gh}%
\end{equation}
The generating equations (\ref{3.3}) with the boundary condition $\left.
S\right\vert _{\phi^{\ast}=\bar{\phi}=0}=S_{0}$ are solved by a functional
linear in the antifields (for details, see (\ref{solxy}), (\ref{xy}) in
Appendix~\ref{AppB})%
\begin{equation}
S=S_{0}+\int d^{D}x\left(  A_{\mu a}^{\ast m}X_{1}^{\mu ma}+B_{a}^{\ast
m}X_{2}^{ma}+C_{ab}^{\ast m}X_{3}^{mab}+\bar{A}_{\mu}^{m}Y_{1}^{\mu m}+\bar
{C}_{a}^{m}Y_{3}^{ma}\right)  \,, \label{solYM}%
\end{equation}
where the functionals $X^{Aa}=\delta S/\delta\phi_{Aa}^{\ast}=\left(
X_{1}^{\mu ma},X_{2}^{ma},X_{3}^{mab}\right)  $ and $Y^{A}=\delta S/\delta
\bar{\phi}_{A}=\left(  Y_{1}^{\mu m},Y_{2}^{m},Y_{3}^{ma}\right)  $ are given
by%
\begin{align}
&  X_{1}^{\mu ma}=D^{\mu mn}C^{na}\ , &  &  Y_{1}^{\mu m}=D^{\mu mn}%
B^{n}+\frac{1}{2}f^{mnl}C^{la}D^{\mu nk}C^{kb}\varepsilon_{ba}\ ,\nonumber\\
&  X_{2}^{ma}=-\frac{1}{2}f^{mnl}B^{l}C^{na}-\frac{1}{12}f^{mnl}f^{lrs}%
C^{sb}C^{ra}C^{nc}\varepsilon_{cb}\ , &  &  Y_{2}^{m}=0\ ,\label{xyYM}\\
&  X_{3}^{mab}=-\varepsilon^{ab}B^{m}-\frac{1}{2}f^{mnl}C^{lb}C^{na}\ , &  &
Y_{3}^{ma}=f^{mnl}B^{l}C^{na}+\frac{1}{6}f^{mnl}f^{lrs}C^{sb}C^{ra}%
C^{nc}\varepsilon_{cb}\ .\nonumber
\end{align}
Hence, the finite BRST-antiBRST transformations $\Delta\phi^{A}=X^{Aa}%
\lambda_{a}-\left(  1/2\right)  Y^{A}\lambda^{2}$ read as follows:%
\begin{align}
\Delta A_{\mu}^{m}  &  =D_{\mu}^{mn}C^{na}\lambda_{a}-\frac{1}{2}\left(
D_{\mu}^{mn}B^{n}+\frac{1}{2}f^{mnl}C^{la}D_{\mu}^{nk}C^{kb}\varepsilon
_{ba}\right)  \lambda^{2}\ ,\label{DAmm}\\
\Delta B^{m}  &  =-\frac{1}{2}\left(  f^{mnl}B^{l}C^{na}+\frac{1}{6}%
f^{mnl}f^{lrs}C^{sb}C^{ra}C^{nc}\varepsilon_{cb}\right)  \lambda
_{a}\ ,\label{DBm}\\
\Delta C^{ma}  &  =\left(  \varepsilon^{ab}B^{m}-\frac{1}{2}f^{mnl}%
C^{la}C^{nb}\right)  \lambda_{b}-\frac{1}{2}\left(  f^{mnl}B^{l}C^{na}%
+\frac{1}{6}f^{mnl}f^{lrs}C^{sb}C^{ra}C^{nc}\varepsilon_{cb}\right)
\lambda^{2}\ , \label{DCma}%
\end{align}
where the approximation linear in $\lambda_{a}=\mu_{a}$ produces the
infinitesimal BRST-antiBRST transformations $\delta\phi^{A}=X^{Aa}\mu
_{a}=\left(  s^{a}\phi^{A}\right)  \mu_{a}$\textbf{.}

To construct the generating functional of Green's functions $Z(J)$ in
(\ref{z(j)}), we choose the gauge functional $F=F\left(  \phi\right)  $ to be
diagonal in $A^{\mu{}m}$, $C^{ma}$, namely,%
\begin{equation}
F\left(  A,C\right)  =-\frac{1}{2}\int d^{D}x\ \left(  \alpha A_{\mu}%
^{m}A^{m\mu}+\beta\varepsilon_{ab}C^{ma}C^{mb}\right)  \ . \label{F(A,C)}%
\end{equation}
The quantum action $S_{F}(\phi)$ corresponding to this gauge-fixing functional
reads (see (\ref{gengF})--(\ref{S(A,B,C)2}) in Appendix \ref{AppB})%
\begin{equation}
S_{F}(A,B,C)=S_{0}\left(  A\right)  +\left(  1/2\right)  s^{a}s_{a}F\left(
A,C\right)  =S_{0}\left(  A\right)  +S_{\mathrm{gf}}\left(  A,B\right)
+S_{\mathrm{gh}}\left(  A,C\right)  +S_{\mathrm{add}}\left(  C\right)  \ ,
\label{S(A,B,C)}%
\end{equation}
where the gauge-fixing term $S_{\mathrm{gf}}$, the ghost term $S_{\mathrm{gh}%
}$, and the interaction term $S_{\mathrm{add}}$, quartic in $C^{ma}$, are
given by%
\begin{align}
S_{\mathrm{gf}}  &  =\int d^{D}x\ \left[  \alpha\left(  \partial^{\mu}A_{\mu
}^{m}\right)  -\beta B^{m}\right]  B^{m}\,,\,\,\,S_{\mathrm{gh}}=\frac{\alpha
}{2}\int d^{D}x\ \left(  \partial^{\mu}C^{ma}\right)  D_{\mu}^{mn}%
C^{nb}\varepsilon_{ab}\ ,\label{Sgh}\\
S_{\mathrm{add}}  &  =\frac{\beta}{24}\int d^{D}x\ \ f^{mnl}f^{lrs}%
C^{sa}C^{rc}C^{nb}C^{md}\varepsilon_{ab}\varepsilon_{cd}\,. \label{Sadd}%
\end{align}

Let us examine the choice of the coefficients $\alpha$, $\beta$ leading to
$R_{\xi}$-like gauges. Namely, in view of the contribution $S_{\mathrm{gf}}$
to the quantum action $S_{F}$,
\begin{equation}
S_{\mathrm{gf}}=\int d^{D}x\ \left[  \alpha\left(  \partial^{\mu}A_{\mu}%
^{m}\right)  -\beta B^{m}\right]  B^{m}\ , \label{Sgfxi}%
\end{equation}
we impose the conditions%
\begin{equation}
\alpha=1\ ,\ \ \ \beta=-\frac{\xi}{2}\ . \label{coeffab}%
\end{equation}
Thus, the gauge-fixing functional $F_{\left(  \xi\right)  }=F_{\left(
\xi\right)  }\left(  A,C\right)  $ corresponding to an $R_{\xi}$-like gauge
can be chosen as%
\begin{align}
F_{\left(  \xi\right)  }  &  =\frac{1}{2}\int d^{D}x\ \left(  -A_{\mu}%
^{m}A^{m\mu}+\frac{\xi}{2}\varepsilon_{ab}C^{ma}C^{mb}\right)
\,,\,\,\,\mathrm{so}\,\,\mathrm{that}\label{Fxi}\\
F_{\left(  0\right)  }  &  =-\frac{1}{2}\int d^{D}x\ A_{\mu}^{m}A^{m\mu}%
\quad\mathrm{and}\quad F_{\left(  1\right)  }\ =\ \frac{1}{2}\int
d^{D}x\ \left(  -A_{\mu}^{m}A^{m\mu}+\frac{1}{2}\varepsilon_{ab}C^{ma}%
C^{mb}\right)  \ , \label{F01}%
\end{align}
where the gauge-fixing functional $F_{\left(  0\right)  }\left(  A\right)  $
induces the contribution $S_{\mathrm{gf}}\left(  A,B\right)  $ to the quantum
action that arises in the case of the Landau gauge $\chi(A)=\partial^{\mu
}A_{\mu}^{m}=0$ for $(\alpha,\beta)=(1,0)$ in (\ref{Sgfxi}), whereas the
functional $F_{\left(  1\right)  }\left(  A,C\right)  $ corresponds to the
Feynman (covariant) gauge $\chi(A,B)=\partial^{\mu}A_{\mu}^{m}+\left(
1/2\right)  B^{m}=0$ for $(\alpha,\beta)=(1,-1/2)$ in (\ref{Sgfxi})

Let us find the parameters $\lambda_{a}=s_{a}\Lambda$ of a finite
field-dependent BRST-antiBRST transformation that connects an $R_{\xi}$ gauge
with an $R_{\xi+\Delta\xi}$ gauge, according to (\ref{2}), where%
\begin{equation}
\Delta F_{\left(  \xi\right)  }\ =\ F_{\left(  \xi+\Delta\xi\right)
}-F_{\left(  \xi\right)  }=\frac{\Delta\xi}{4}\varepsilon_{ab}\int
d^{D}x\ C^{ma}C^{mb}\ . \label{DFxi}%
\end{equation}
Explicitly,%
\begin{equation}
\delta\left(  \Delta F_{\left(  \xi\right)  }\right)  =s^{a}\left(  \Delta
F_{\left(  \xi\right)  }\right)  \mu_{a}=\frac{\Delta\xi}{2}\varepsilon
_{ba}\int d^{D}x\ C^{mb}\delta C^{ma}\ , \label{dDFxi}%
\end{equation}
where $\delta C^{ma}=\left(  \varepsilon^{ab}B^{m}-\left(  1/2\right)
f^{mnl}C^{la}C^{nb}\right)  \mu_{b}$ is the linear part of the finite
BRST-antiBRST transformation (\ref{DCma}), which implies%
\begin{equation}
s^{a}\left(  \Delta F_{\left(  \xi\right)  }\right)  \ =\ \frac{\Delta\xi}%
{2}\varepsilon_{bc}\int d^{D}x\ C^{mb}\left(  \varepsilon^{ca}B^{m}-\frac
{1}{2}f^{mnl}C^{lc}C^{na}\right)  \ . \label{saDF}%
\end{equation}
In order to calculate $s^{a}s_{a}\left(  \Delta F_{\left(  \xi\right)
}\right)  $, we remind that%
\begin{align}
\frac{1}{2}s^{a}s_{a}F_{\left(  \xi\right)  }  &  =\left.  S_{\mathrm{gf}%
}+S_{\mathrm{gh}}+S_{\mathrm{add}}\right\vert _{\alpha=1,\beta=-\xi
/2}\nonumber\\
&  =\int d^{D}x\ \left\{  \left[  \left(  \partial^{\mu}A_{\mu}^{m}\right)
+\frac{\xi}{2}B^{m}\right]  B^{m}+\frac{1}{2}\left(  \partial^{\mu}%
C^{ma}\right)  D_{\mu}^{mn}C^{nb}\varepsilon_{ab}-\frac{\xi}{48}%
\ f^{mnl}f^{lrs}C^{sa}C^{rc}C^{nb}C^{md}\varepsilon_{ab}\varepsilon
_{cd}\right\}  \ , \label{s2Fxi}%
\end{align}
whence
\begin{equation}
s^{a}s_{a}\left(  \Delta F_{\left(  \xi\right)  }\right)  \ =\ \Delta\xi\int
d^{D}x\ \left(  B^{m}B^{m}-\frac{1}{24}\ f^{mnl}f^{lrs}C^{sa}C^{rc}%
C^{nb}C^{md}\varepsilon_{ab}\varepsilon_{cd}\right)  \ . \label{s2FDxi}%
\end{equation}
Finally, the functionals $\lambda_{a}\left(  \phi\right)  $ that connect an
$R_{\xi}$-like gauge to an $R_{\xi+\Delta\xi}$-like gauge are given by
(\ref{2})%
\begin{align}
\lambda_{a} =\frac{\Delta\xi}{4i\hbar}\varepsilon_{ab}\int d^{D}x\ \left(
B^{n}C^{nb} \right)  \sum_{n=0}^{\infty}\frac{1}{\left(  n+1\right)  !}\left[
\frac{1}{4i\hbar}\Delta\xi\int d^{D}y\ \left(  B^{u}B^{u}-\frac{1}%
{24}\ f^{uwt}f^{trs}C^{sc}C^{rp}C^{wd}C^{uq}\varepsilon_{cd}\varepsilon
_{pq}\right)  \right]  ^{n}\ . \label{lamaxi}%
\end{align}
In particular, the first order of $\lambda_{a}=\mu_{a}$ in powers of $\Delta
F_{\left(  \xi\right)  }$ has the form (\ref{linear})%
\begin{equation}
\mu_{a}\ =\ -\frac{i}{2\hbar}s_{a}\Delta F_{\left(  \xi\right)  }%
=-\frac{i\Delta\xi}{4\hbar}\varepsilon_{ab}\int d^{D}x\ B^{m}C^{mb} \ .
\label{linlaDxi}%
\end{equation}

We have thus solved the problem of reaching any gauge in the family of
$R_{\xi}$-like gauges, starting from a certain gauge encoded in the path
integral by a functional $F_{\left(  \xi\right)  }$, within the framework of
BRST-antiBRST quantization for Yang--Mills theories by means of finite
BRST-antiBRST transformations with field-dependent parameters $\lambda_{a}$ in
(\ref{lamaxi}). Generally, if the BRST-antiBRST invariant quantum action
$S_{F_{0}}$ of a Yang--Mills theory is given in terms of a gauge induced by a
gauge-fixing functional $F_{0}$, then, in order to reach the quantum action
$S_{F}$ in terms of another gauge induced by a gauge-fixing functional $F$, it
is sufficient to make a change of variables in the path integral (\ref{z(j)})
with $S_{F_{0}}$, given by a finite field-dependent BRST-antiBRST
transformation with an $\mathrm{Sp}(2)$-doublet of the odd-valued functionals
\begin{align}
\label{lamagen}\lambda_{a}(F-F_{0})=\frac{1}{2i\hbar}\left[  s_{a}(F-F_{0})
\right]  \sum_{n=0}^{\infty}\frac{1}{\left(  n+1\right)  !}\left(  \frac
{1}{4i\hbar}s^{b}s_{b}(F-F_{0}) \right)  ^{n}.
\end{align}
In particular, if we choose $F_{0}=F_{\left(  \xi\right)  }$, with $F_{\left(
\xi\right)  }$ given by (\ref{Fxi}), then the above relation (\ref{lamagen})
describes the transition from an $R_{\xi}$-like gauge to a gauge parameterized
by an arbitrary gauge-fixing functional $F=F\left(  A,B,C\right)  $.

\section{Gribov--Zwanziger Action in $R_{\xi}$-like Gauges}

\label{GZ} \setcounter{equation}{0}

Let us extend the construction of the Gribov horizon \cite{Gribov} to the case
of a BRST-antiBRST invariant Yang--Mills theory in a way consistent with the
gauge-independence of the $S$-matrix. To this end, we examine the sum of the
Yang--Mills quantum action (\ref{S(A,B,C)}) in the Landau gauge $\partial
^{\mu}A_{\mu}^{m}=0$ (with the gauge-fixing functional $F_{(0)}$ in
(\ref{F01}) corresponding to the case $\alpha=1$, $\beta=0$) and the non-local
horizon functional \cite{Zwanziger}%
\begin{equation}
\label{gribovH}h\left(  A\right)  =\gamma^{2}\int d^{D}x \left(\int d^{D}y\ \ f^{mrl}%
A_{\mu}^{r}\left(  x\right)  \left(  K^{-1}\right)  ^{mn}\left(  x;y\right)
f^{nsl}A^{\mu{}s }\left(  y\right)  +D\left(  N^{2}-1\right) \right) \ .
\end{equation}
where $K^{-1}$ is the inverse,%
\begin{equation}
\label{K^{-1}}\int d^{D}z\ \left(  K^{-1}\right)  ^{ml}\left(  x;z\right)
\left(  K\right)  ^{ln}\left(  z;y\right)  =\int d^{D}z\ \left(
K^{-1}\right)  ^{nl}\left(  x;z\right)  \left(  K\right)  ^{lm}\left(
z;y\right)  =\delta^{mn}\delta\left(  x-y\right)  \ ,
\end{equation}
of the Faddeev--Popov operator $K$ induced by the gauge-fixing functional
$F_{(\xi\to0)}$ corresponding to the Landau gauge $\partial^{\mu}A_{\mu}^{m}=0
$ in the BRST approach,%
\begin{equation}
\label{K}K^{mn}\left(  x;y\right)  =\left(  \delta^{mn}\partial^{2}%
+f^{mln}A_{\mu}^{l}\partial^{\mu}\right)  \delta\left(  x-y\right)
\ ,\ \ \ K^{mn}\left(  x;y\right)  =K^{nm}\left(  y;x\right)  \,,
\end{equation}
whereas $\gamma\in\mathbb{R}$ is the so-called thermodynamic, or Gribov,
parameter \cite{Zwanziger}, introduced in a self-consistent way by the gap
equation for an analogue $S_{h}$ of the Gribov--Zwanziger action in the
BRST-antiBRST approach:
\begin{equation}
\frac{\partial}{\partial\gamma}\left\{  \frac{\hbar}{i}\,\ln\,\left[
\int\!D\phi\ \exp\left(  \frac{i}{\hbar}S_{h}\right)  \right]  \right\}
=\frac{\partial\mathcal{E}_{\mathrm{vac}}}{\partial\gamma}\ =0\ .
\label{gapeq}%
\end{equation}
In (\ref{gapeq}), we have used the definition of the vacuum energy
$\mathcal{E}_{\mathrm{vac}}$ and introduced a modified quantum action for the
Gribov--Zwanziger model as an additive extension of the Yang--Mills quantum
action $S_{F_{0}}$ (\ref{S(A,B,C)}) in the Landau gauge:
\begin{equation}
\label{GZbab}S_{h}\left(  \phi\right)  =S_{F_{0}}\left(  \phi\right)
+h\left(  \phi\right)  \,,\,\,\,F_{0}=F_{\left(  0\right)  }\,,
\end{equation}
The action $S_{h}\left(  \phi\right)  $ is not invariant under the finite
BRST-antiBRST transformations:%
\begin{equation}
\label{nbabh}\Delta S_{h}=\Delta h=\left(  s^{a}h\right)  \lambda_{a}+\frac
{1}{4}\left(  s^{2}h\right)  \lambda^{2}\ne0\ ,
\end{equation}
indeed, according to $\Delta\phi^{A}=\left(  s^{a}\phi^{A}\right)  \lambda
_{a}+\left(  1/4\right)  \left(  s^{2}\phi^{A}\right)  \lambda^{2}$, with
allowance for (\ref{DAmm})--(\ref{DCma}), (\ref{s2(AB)}), we have%
\begin{align}
s^{a}h  &  =\gamma^{2}f^{mrk}f^{kns}\int d^{D}x\ d^{D}y\ \left[  2D_{\mu}%
^{rl}C^{la}\left(  x\right)  \left(  K^{-1}\right)  ^{mn}\left(  x;y\right)
\right. \nonumber\\
&  -f^{utv}\int d^{D}x^{\prime}\ d^{D}y^{\prime}\left.  A_{\mu}^{r}{\left(
x\right)  }\left(  K^{-1}\right)  ^{mu}\left(  x;x^{\prime}\right)
K^{tl}\left(  x^{\prime};y^{\prime}\right)  C^{la}\left(  y^{\prime}\right)
\left(  K^{-1}\right)  ^{vn}\left(  y^{\prime};y\right)  \right]  A^{s\mu
}\left(  y\right)  \label{sah}%
\end{align}
and
\begin{align}
s^{2}h  &  =\gamma^{2}\ f^{mrk}f^{kns}\int d^{D}x\ d^{D}y\ \left\{  4\left(
-D_{\mu}^{rt}B^{t}+\frac{1}{2}f^{rtl}C^{la}D_{\mu}^{tu}C^{ub}\varepsilon
_{ab}\right)  \left(  x\right)  \left(  K^{-1}\right)  ^{mn}\left(
x;y\right)  A^{s\mu}\left(  y\right)  \right. \nonumber\\
&  +2\varepsilon_{ab}D_{\mu}^{rl}C^{la}\left(  x\right)  \left(
K^{-1}\right)  ^{mn}\left(  x;y\right)  D^{st\mu}C^{tb}\left(  y\right)
\nonumber\\
&  -4\varepsilon_{ab}f^{utv}\int d^{D}x^{\prime}\ d^{D}y^{\prime}\ D_{\mu
}^{rl}C^{la}\left(  x\right)  \left(  K^{-1}\right)  ^{mu}\left(  x;x^{\prime
}\right)  K^{tw}\left(  x^{\prime};y^{\prime}\right)  C^{wb}\left(  y^{\prime
}\right)  \left(  K^{-1}\right)  ^{vn}\left(  y^{\prime};y\right)  A^{s\mu
}\left(  y\right) \nonumber\\
&  +f^{utv}\int d^{D}x^{\prime}\ d^{D}y^{\prime}\ A_{\mu}^{r}\left(  x\right)
\left[  -\varepsilon_{ab}f^{u^{\prime}t^{\prime}v^{\prime}}\int d^{D}%
x^{\prime\prime}\ d^{D}y^{\prime\prime}\ \left(  K^{-1}\right)  ^{mu^{\prime}%
}\left(  x;x^{\prime\prime}\right)  K^{t^{\prime}l^{\prime}}\left(
x^{\prime\prime};y^{\prime\prime}\right)  C^{l^{\prime}a}\left(
y^{\prime\prime}\right)  \right. \nonumber\\
&  \times\left(  K^{-1}\right)  ^{v^{\prime}u}\left(  y^{\prime\prime
};x^{\prime}\right)  K^{tl}\left(  x^{\prime};y^{\prime}\right)  C^{lb}\left(
y^{\prime}\right)  \left(  K^{-1}\right)  ^{vn}\left(  y^{\prime};y\right)
-\varepsilon_{ab}f^{tlt^{\prime}}\left(  K^{-1}\right)  ^{mu}\left(
x;x^{\prime}\right)  K^{t^{\prime}l^{\prime}}\left(  x^{\prime};y^{\prime
}\right) \nonumber\\
&  \times C^{l^{\prime}a}(y^{\prime})C^{lb}\left(  x^{\prime}\right)  \left(
K^{-1}\right)  ^{vn}\left(  y^{\prime};y\right)  +2\left(  K^{-1}\right)
^{mu}\left(  x;x^{\prime}\right)  K^{tl}\left(  x^{\prime};y^{\prime}\right)
B^{l}\left(  y^{\prime}\right)  \left(  K^{-1}\right)  ^{vn}\left(  y^{\prime
};y\right) \nonumber\\
&  +\varepsilon_{ab}f^{u^{\prime}t^{\prime}v^{\prime}}\left(  K^{-1}\right)
^{mu}\left(  x;x^{\prime}\right)  K^{tl}\left(  x^{\prime};y^{\prime}\right)
C^{la}\left(  y^{\prime}\right) \nonumber\\
&  \times\left.  \left.  \int d^{D}x^{\prime\prime}\ d^{D}y^{\prime\prime
}\ \left(  K^{-1}\right)  ^{vu^{\prime}}\left(  y^{\prime};x^{\prime\prime
}\right)  K^{t^{\prime}l^{\prime}}\left(  x^{\prime\prime};y^{\prime\prime
}\right)  C^{l^{\prime}b}\left(  y^{\prime\prime}\right)  \left(
K^{-1}\right)  ^{v^{\prime}n}\left(  y^{\prime\prime};y\right)  \right]
A^{s\mu}\left(  y\right)  \right\}  \, , \label{s2h}%
\end{align}
where we have used the identity
\begin{equation}
\label{auxsaK}s^{a} K^{mn}\left(  x;y\right)  \ =\ f^{mrn}K^{rs}\left(
x;y\right)  C^{sa}(y)\,.
\end{equation}

To determine the horizon functional for a general $R_{\xi}$-like gauge in the
BRST-antiBRST description, we propose
\begin{align}
h_{\xi}  &  =h+\frac{1}{2i\hbar}\left(  s^{a}h\right)  \left(  s_{a}\Delta
F_{\left(  \xi\right)  }\right)  \sum_{n=0}^{\infty}\frac{1}{\left(
n+1\right)  !}\left(  \frac{1}{4i\hbar}s^{b}s_{b}\Delta F_{\left(  \xi\right)
}\right)  ^{n}\nonumber\\
&  -\frac{1}{16\hbar^{2}}\left(  s^{2}h\right)  \left(  s\Delta F_{\left(
\xi\right)  }\right)  ^{2}\left[  \sum_{n=0}^{\infty}\frac{1}{\left(
n+1\right)  !}\left(  \frac{1}{4i\hbar}s^{b}s_{b}\Delta F_{\left(  \xi\right)
}\right)  ^{n}\right]  ^{2}\ . \label{gribovHxi}%
\end{align}
Here, $s^{a}h$ and $s^{2}h$ are given by (\ref{sah}), (\ref{s2h}), while
$s_{a}\Delta F_{\left(  \xi\right)  }$ and $s^{a}s_{a}\Delta F_{\left(
\xi\right)  }$ are given by (\ref{saDF}), (\ref{s2FDxi}) for $\Delta\xi=\xi$,
whereas the $\mathrm{Sp}(2)$-doublet $\lambda_{\xi}^{a}(\phi)$ of
field-dependent anticommuting parameters in (\ref{lamaxi}) relates the Landau
gauge to an arbitrary $R_{\xi}$-like gauge:
\begin{align}
\Delta F_{\left(  \xi\right)  }  &  =F_{\left(  \xi\right)  }-F_{\left(
0\right)  }=\frac{\xi}{4}\varepsilon_{ab}\int d^{D}x\ C^{ma}C^{mb}%
,\label{DFxin}\\
s_{a}\Delta F_{\left(  \xi\right)  }  &  =\frac{\xi}{2}\varepsilon_{ab}\int
d^{D}x\ B^{m}C^{mb} ,\label{saDFxin2}\\
s^{a}s_{a}\Delta F_{\left(  \xi\right)  }  &  =\xi\int d^{D}x\ \left(
B^{m}B^{m}-\frac{1}{24}\ f^{mnl}f^{lrs}C^{sa}C^{rc}C^{nb}C^{md}\varepsilon
_{ab}\varepsilon_{cd}\right)  \,. \label{saDFxin3}%
\end{align}
In particular, the approximation linear in $\xi$ implies, $\lambda_{\xi}%
^{a}(\phi)=s^{a}\Lambda_{\xi}(\phi)$ for $\Lambda_{\xi}(\phi)=\frac{\xi
}{8i\hbar}\varepsilon_{ab}\int d^{D}x\ C^{ma}C^{mb}$,
\begin{align}
h_{\xi}\left(  \phi\right)   &  =h\left(  A\right)  +\frac{\xi}{4i\hbar
}\varepsilon_{ab}\gamma^{2}f^{mrl}f^{lns}\int d^{D}x\ d^{D}y\ \left[  2D_{\mu
}^{rk}C^{ka}\left(  x\right)  \left(  K^{-1}\right)  ^{mn}\left(  x;y\right)
-f^{m^{\prime}l^{\prime}n^{\prime}}\int d^{D}x^{\prime}\ d^{D}y^{\prime
}\ A_{\mu}^{r}\left(  x\right)  \right. \nonumber\\
&  \phantom{\int}\times\left.  \left(  K^{-1}\right)  ^{mm^{\prime}}\left(
x;x^{\prime}\right)  K^{l^{\prime}t^{\prime}}\left(  x^{\prime};y^{\prime
}\right)  C^{t^{\prime}a}\left(  y^{\prime}\right)  \left(  K^{-1}\right)
^{n^{\prime}n}\left(  y^{\prime};y\right)  \right]  A^{s\mu}\left(  y\right)
\int d^{D}z\left(  B^{w}C^{wb} \right)  \,.\label{hxilin}%
\end{align}
Notice that even the approximation to $h_{\xi}\left(  \phi\right)  $ being
linear in powers of $\xi$ is different from the proposal \cite{LL2} for the
horizon functional given by $R_{\xi}$-gauges in terms of field-dependent BRST
transformations, which reflects the $\mathrm{Sp}(2)$-symmetric character of
the dependence of $h_{\xi}\left(  \phi\right) $ on the ghost and antighost
fields $C^{ma}$.

The proposal (\ref{gribovHxi}) for the Gribov horizon functional in a general
$R_{\xi}$-gauge is consistent with the study of gauge-independence for the
generating functional of Green's functions, determined for a BRST-antiBRST
extension of the Gribov--Zwanziger model as follows:
\begin{equation}
Z_{\mathrm{GZ},F_{0}}(J)=\int d\phi\ \exp\left\{  \frac{i}{\hbar}\left[
S_{h}\left(  \phi\right)  +J_{A}\phi^{A}\right]  \right\}  \;. \label{z(j)GZ}%
\end{equation}
Indeed, making in the path integral for $Z_{\mathrm{GZ},F_{0}}(J)$ a change of
variables being a finite field-dependent BRST-antiBRST transformation with the
parameters $\lambda_{\xi}^{a}(\phi)$ given by (\ref{lamaxi}), where $\Delta
\xi=\xi$, we find, due to the fact that the Yang--Mills quantum action
$S_{F_{0}}(\phi)$ transforms to $S_{F_{\xi}}(\phi)$, with $F_{\xi}=F_{\left(
\xi\right)  }$,
\begin{equation}
Z_{\mathrm{GZ},F_{0}}(J)=\int d\phi\ \exp\left\{  \frac{i}{\hbar}\left[
S_{F_{\xi}}(\phi)+h_{\xi}\left(  \phi\right)  +J_{A}\phi^{A}+J_{A}\Delta
\phi^{A}\right]  \right\}  \;, \label{z(j)GZ2}%
\end{equation}
where $h_{\xi}\left(  \phi\right)  $ in (\ref{gribovHxi}) corresponds to an
$R_{\xi}$-gauge. As a result, we have
\begin{align}
Z_{\mathrm{GZ},F_{0}}(J)  &  =Z_{\mathrm{GZ},F_{\xi}}(J)\left[  1+\frac
{i}{\hbar}J_{A}\left\langle (s^{a}\phi^{A})s_{a}\Lambda(\Delta F_{\left(
\xi\right)  })\right\rangle _{F_{0},J}\right. \nonumber\\
&  +\frac{i}{4\hbar}\left.  J_{A}\left\langle (s^{2}\phi^{A})\left[
s\Lambda(\Delta F_{\left(  \xi\right)  })\right]  ^{2}+\frac{i}{\hbar
}\varepsilon_{ab}(s^{a}\phi^{A})J_{B}(s^{b}\phi^{B})\left[  s\Lambda(\Delta
F_{\left(  \xi\right)  })\right]  ^{2}\right\rangle _{F_{0},J}\right]  \;,
\label{z(j)GZv}%
\end{align}
where the vacuum expectation value is computed with respect to $Z_{\mathrm{GZ}%
,F}(J)$. The relation (\ref{z(j)GZv}) implies that neither the functional
$Z_{\mathrm{GZ},F_{\xi}}(J)$ nor the $S$-matrix depends on the gauge
(parameter $\xi$) at the extremals given by $J_{A}=0$. This justifies our
proposal for the horizon functional in the form\footnote{There exist other
ways to obtain the Gribov horizon functional $h_{\xi}$ for gauges beyond the
Landau gauge, see, e.g., \cite{sorellas,Gongyo}; however, in view of its
non-pertubative character \cite{Zwanziger}, the derivation procedure faces the
problem of gauge dependence.} (\ref{gribovHxi}). At the same time, we note
that the Gribov--Zwanziger model in BRST-antiBRST quantization encounters the
problem of unitarity, since the gauge degrees of freedom, being non-dynamical
in the Yang--Mills theory, should now be regarded as dynamical ones, due to
the explicit form of the horizon functional $h_{\xi}\left(  \phi\right)  $.

Finally, it is possible to construct a Gribov horizon functional $h_{F}(\phi)$
in any differential gauge\footnote{Due to the result of Singer \cite{Singer},
Gribov copies should arise in non-Abelian gauge theories in case a
differential gauge is used to fix the gauge ambiguity.} induced by a
gauge-fixing functional $F\left(  \phi\right)  $, starting from the horizon
functional $h(A)$ in the Landau gauge, corresponding to the gauge-fixing
functional $F_{0}(A)$. To this end, it is sufficient to make a change of
variables in the path integral (\ref{z(j)GZ}), given by a finite
field-dependent BRST-antiBRST transformation with the $\mathrm{Sp}(2)$-doublet
$\lambda_{a}(F-F_{0})$ of odd-valued functionals given by (\ref{lamagen}).
Thus, the functional $h_{F}(\phi)$ reads as follows:
\begin{align}
h_{F}  &  =h+\frac{1}{2i\hbar}\left(  s^{a}h\right)  \left[  s_{a}%
(F-F_{0})\right]  \sum_{n=0}^{\infty}\frac{1}{\left(  n+1\right)  !}\left(
\frac{1}{4i\hbar}s^{b}s_{b}(F-F_{0})\right)  ^{n}\nonumber\\
&  -\frac{1}{16\hbar^{2}}\left(  s^{2}h\right)  \left[  s(F-F_{0})\right]
^{2}\left[  \sum_{n=0}^{\infty}\frac{1}{\left(  n+1\right)  !}\left(  \frac
{1}{4i\hbar}s^{b}s_{b}(F-F_{0})\right)  ^{n}\right]  ^{2}\,.
\label{gribovHgen}%
\end{align}
Generally, a finite change $F\rightarrow F+\Delta F$ of the gauge condition
induces a finite change of any functional $G_{F}(\phi)$, so that in the
reference frame corresponding to the gauge $F+\Delta F$ it can be represented
according to (\ref{DeltaF}), (\ref{lamagen}),
\begin{equation}
G_{F+\Delta F}=G_{F}+\left(  s^{a}G_{F}\right)  \lambda_{a}\left(  \Delta
F\right)  +\frac{1}{4}\left(  s^{2}G_{F}\right)  \lambda_{a}\left(  \Delta
F\right)  \lambda^{a}\left(  \Delta F\right)  \,, \label{finvarfun}%
\end{equation}
which is an extension of the infinitesimal change $G_{F}\rightarrow
G_{F}+\delta G_{F}$ induced by a variation of the gauge, $F\rightarrow
F+\delta F$,
\begin{equation}
G_{F+\delta F}=G_{F}-\frac{i}{2\hbar}\left(  s^{a}G_{F}\right)  \left(
s_{a}\delta F\right)  \,, \label{infinvarfungt}%
\end{equation}
corresponding, in the case $G_{F}\left(  A\right)  $, to the gauge
transformations (\ref{R(A)}), with the functions $\zeta^{m}(x)$ given below
\begin{equation}
\delta G_{F}=G_{F+\delta F}-G_{F}=\int d^{D}x\frac{\delta G_{F}}{\delta A^{\mu
m}(x)}D^{mn\mu}\zeta^{n}(x)\,,\,\,\,\mathrm{\ where}\,\,\,\zeta^{m}%
(x)=-\frac{i}{2\hbar}C^{ma}(x){\left(  s_{a}\delta F\right)  }\,.
\label{infinvarfun}%
\end{equation}
Due to the presence of the term with $s^{2}G_{F}$ in a finite gauge variation
of a functional $G_{F}(A)$ depending only on the classical fields $A^{m\mu}$,
the representation (\ref{finvarfun}) is more general than the one that would
correspond to the usual Lagrangian BRST approach (see relation (17) in
\cite{Reshetnyak2}), having the form similar to (\ref{infinvarfun}), and thus
also to (\ref{infinvarfungt}).

We emphasize that the suggested method of using the finite field-dependent
BRST-antiBRST transformations with the purpose of finding the
Gribov--Zwanziger horizon functional in any differential gauge, starting from
the Gribov--Zwanziger theory in the Landau gauge, is valid in perturbation
theory and preserves the number of physical degrees of freedom, without
entering into contradiction with the result of \cite{Upadhyay2} in the BRST
setting of the problem. However, it is impossible to solve this problem (in
particular, in the Yang--Mills theory) in terms of finite field-dependent
BRST-antiBRST transformations \cite{Upadhyay3}, in view of the absence of a
term being quadratic in powers of the odd-valued parameters, since the
corresponding Yang--Mills quantum action fails to be BRST-antiBRST invariant,
and the Jacobian of the corresponding change of variables with odd-valued
functionally-dependent parameters does not generate terms which are entirely
BRST-antiBRST-exact. These terms change the BRST-antiBRST-exact part of the
action, as well as the extremals; however, they do not affect the number of
physical degrees of freedom.

\section{Discussion}

\label{Concl} \setcounter{equation}{0}

In the present work, we have proposed the concept of finite
BRST-antiBRST transformations for Yang--Mills theories in the
$\mathrm{Sp}(2)$-covariant Lagrangian quantization \cite{BLT1,
BLT2}, realized in the form (\ref{finBRSTantiBRST}),
(\ref{finite}), being polynomial in powers of a constant
$\mathrm{Sp}\left(  2\right)  $-doublet of anticommuting Grassmann
parameters $\lambda_{a}$ and leaving the quantum action of the
Yang--Mills theory invariant to all orders in $\lambda_{a}$. In
the case of constant $\lambda_{a}$, the set of finite
BRST-antiBRST transformations forms an Abelian two-parametric Lie
supergroup with the elements $g(\lambda)=\exp\left(
\overleftarrow{s}^{a}\lambda_{a}\right)  =\left(
1+\overleftarrow{s}^{a}\lambda_{a}+\frac{1}{4}\overleftarrow{s}^{a}%
\overleftarrow{s}_{a}\lambda^{2}\right)  $, so that
$\Delta\phi^{A}=\phi ^{A}\left[  \exp\left(
\overleftarrow{s}^{a}\lambda_{a}\right)  -1\right]  $, where
$G\overleftarrow{s}^{a}\equiv s^{a}G$, for any $G=G\left(\phi
\right)$. Secondly, this ensures exact invariance of the integrand
in the generating functional of Green's functions $Z_{F}(J)$ with
vanishing external sources $J_{A}$ and also allows one to obtain
the Ward identities.

We have determined the finite field-dependent BRST-antiBRST transformations as
polynomials in the $\mathrm{Sp}\left(  2\right)  $-doublet of Grassmann-odd
functionals $\lambda_{a}(\phi)$, depending on the whole set of fields that
compose the configuration space of Yang--Mills theories, and have also
calculated the Jacobian (\ref{superJ1}) corresponding to this change of
variables by using a special class of transformations with $s_{a}$-potential
parameters $\lambda_{a}(\phi)=s_{a}\Lambda(\phi)$ for a Grassmann-even
functional $\Lambda(\phi)$ and Grassmann-odd generators $s_{a}$ of
BRST-antiBRST transformations.

In comparison with finite field-dependent BRST transformations in Yang--Mills
theories \cite{LL1}, in which a change of the gauge corresponds to a unique
field-dependent parameter (up to BRST-exact terms), it is only
functionally-dependent finite field-dependent BRST-antiBRST transformations
with $\lambda_{a} = s_{a}\Lambda(\Delta F)$ that are in one-to-one
correspondence with $\Delta F$. We have found (\ref{Lambda-Fsol1}) a solution
$\Lambda(\Delta F)$ to the so-called compensation equation (\ref{Lambda-F})
for an unknown functional $\Lambda$ generating an $\mathrm{Sp}\left(
2\right)  $-doublet $\lambda_{a}$, in order to establish a relation of the
Yang--Mills quantum action $S_{F}$ in a certain gauge determined by a gauge
Boson $F$ with the action $S_{F+\Delta F}$ induced by a different gauge
$F+\Delta F$. This makes it possible to investigate the problem of
gauge-dependence for the generating functional $Z_{F}(J)$ under a finite
change of the gauge in the form (\ref{gaugedepzj}), leading to the
gauge-independence of the physical $S$-matrix.

In terms of the potential $\Lambda$ inducing the finite field-dependent
BRST-antiBRST transformations, we have explicitly constructed (\ref{lamaxi})
the parameters $\lambda_{a}$ generating a change of the gauge in the path
integral for Yang--Mills theories within a class of linear $R_{\xi}$-like
gauges related to even-valued gauge-fixing functionals $F_{\left(  \xi\right)
}$, with $\xi=0,1$ corresponding to the Landau and Feynman (covariant) gauges,
respectively. We have shown how to reach an arbitrary gauge given by a gauge
Boson $F$ within the path integral representation, starting from the reference
frame with a gauge Boson $F_{0}$ by means of finite field-dependent
BRST-antiBRST transformations with the parameters $\lambda_{a}(F-F_{0})$ given
by (\ref{lamagen}).

We have applied the concept of finite field-dependent BRST-antiBRST
transformations to construct the Gribov horizon functional $h_{\xi}$, given by
(\ref{gribovHxi}) in arbitrary $R_{\xi}$-like gauges, starting from a
previously known BRST-antiBRST non-invariant functional $h$, as in
\cite{Zwanziger}, corresponding to the Landau gauge and induced by an
even-valued functional $F_{\left(  0\right)  }$. The construction is
consistent with the study of gauge-independence for the generating functionals
of Green's functions $Z_{\mathrm{GZ},F_{0}}(J)$ in (\ref{z(j)GZ}) within the
suggested Gribov--Zwanziger model considered in the BRST-antiBRST approach
(\ref{GZbab}).

There are various lines of research for extending the results obtained in the
present work. First, the study of finite field-dependent BRST-antiBRST
transformations for a general gauge theory in the framework of the path
integral\footnote{We have solved this problem in our recent works
\cite{MRnew2, MRnew3}.} (\ref{3.6}). Second, the development of finite
field-dependent BRST transformations for a general gauge theory in the BV
quantization method\footnote{Shortly after the publication of the present
work, we have become aware of the more recent study \cite{BLTfin} of finite
BRST transformations in the BV formalism.} \cite{BV}. Third, the construction
of finite field-dependent BRST-antiBRST transformations in the $\mathrm{Sp}%
(2)$-covariant generalized Hamiltonian quantization \cite{BLT1h,
BLT2h} and the study of their properties in connection with the
corresponding gauge-fixing problem.\footnote{We have solved this
problem in detail \cite{MRnew1}, including the case of Yang--Mills
theories.\label{HamMR}} Fourth, the consideration of the so-called
refined Gribov--Zwanziger theory \cite{0806.0348} in a
BRST-antiBRST setting analogous to \cite{Reshetnyak}, and also the
elaboration of a composite operator technique in the BRST-antiBRST
Lagrangian quantization scheme, in order to examine the Gribov
horizon functional as a composite operator with an external
source, along the lines of \cite{Reshetnyak2}. We also mention the
search for an equivalent local description of the Gribov horizon
functional with a set of auxiliary set fields as in
\cite{Zwanziger} such that it should be consistent with both the
infinitesimal and finite BRST-antiBRST invariance. We are also
interested in the study of the influence of Jacobians generated by
finite field-dependent BRST-antiBRST transformations (linear and
functionally-independent parameters) on the structure of
transformed quantum actions and partition functions \cite{MRnew4}.

Finally, the suggested Gribov horizon functionals beyond the Landau gauge
allow one to study such quantum properties as renormalizability and
confinement within the BRST-antiBRST extension of the Gribov--Zwanziger theory
in a way consistent with the gauge independence of the physical $S$-matrix. We
intend to study these problems in our forthcoming works.

Concluding, let us outline an ansatz for finite field-dependent BRST-antiBRST
transformations of the path integral (\ref{3.6}), corresponding to the case of
a general gauge theory. To this end, notice that the construction
(\ref{finBRSTantiBRST}), (\ref{finite}) of finite BRST-antiBRST
transformations in Section \ref{fBRSTa}, in fact, applies to any infinitesimal
symmetry transformations $\delta\phi^{A}=X^{Aa}\mu_{a}=\left(  s^{a}\phi
^{A}\right)  \mu_{a}$, with anticommuting parameters $\mu_{a}$, $a=1,2$, for a
certain functional $S_{F}\left(  \phi\right)  $, such that $\delta
S_{F}\left(  \phi\right)  =0$, and does not involve any subsidiary conditions
on $X^{Aa}$ and the corresponding $s^{a}$, since the construction is achieved
only by using $Y^{A}=\left(  1/2\right)  X_{,B}^{Aa}X^{Bb}\varepsilon_{ba}$ in
(\ref{finite}), according to (\ref{mixed}). Let us apply this to the vacuum
functional $Z(0)$ of a general gauge theory, given by the path integral
(\ref{3.6}) in the extended space $\Gamma^{p}=\left(  \phi^{A},\phi_{Aa}%
^{\ast},\bar{\phi}_{A},\pi^{Aa},\lambda^{A}\right)  $,%
\begin{equation}
Z(0)=\int d\Gamma\;\exp\left[  \left(  i/\hbar\right)  \mathcal{S}_{F}\left(
\Gamma\right)  \right]  \,\ ,\ \ \ \mathcal{S}_{F}=S+\phi_{Aa}^{\ast}\pi
^{Aa}+\left(  \bar{\phi}_{A}-F_{,A}\right)  \lambda^{A}-\left(  1/2\right)
\varepsilon_{ab}\pi^{Aa}F_{,AB}\pi^{Bb}\ , \label{z(0)}%
\end{equation}
where the integrand $\mathcal{I}_{\Gamma}^{\left(  F\right)  }=d\Gamma
\exp\left[  \left(  i/\hbar\right)  \mathcal{S}_{F}\left(  \Gamma\right)
\right]  $ is invariant, $\delta\mathcal{I}_{\Gamma}^{\left(  F\right)  }=0$,
under the global infinitesimal BRST-antiBRST transformations (\ref{3.7}),
$\delta\Gamma^{p}=\left(  \sigma^{a}\Gamma^{p}\right)  \mu_{a}$, with the
corresponding generators $\sigma^{a}$,%
\begin{equation}
\delta\Gamma^{p}=\left(  \sigma^{a}\Gamma^{p}\right)  \mu_{a}=\delta\left(
\phi^{A},\ \phi_{Ab}^{\ast},\ \bar{\phi}_{A},\ \pi^{Ab},\ \lambda^{A}\right)
=\left(  \pi^{Aa},\ \delta_{b}^{a}S_{,A}\left(  -1\right)  ^{\varepsilon_{A}%
},\ \varepsilon^{ab}\phi_{Ab}^{\ast}\left(  -1\right)  ^{\varepsilon_{A}%
+1},\ \varepsilon^{ab}\lambda^{A},\ 0\right)  \mu_{a}\ . \label{Gamma}%
\end{equation}
In this connection, let us determine finite BRST-antiBRST transformations,
$\Gamma^{p}\rightarrow\Gamma^{p}+\Delta\Gamma^{p}$, parameterized by
anticommuting parameters $\lambda_{a}$, $a=1,2$, as follows:%
\begin{equation}
\mathcal{I}_{\Gamma+\Delta\Gamma}^{\left(  {F}\right)  }=\mathcal{I}_{\Gamma
}^{\left(  {F}\right)  }\ ,\ \ \ \left[  \frac{\overleftarrow{\partial}%
}{\partial\lambda_{a}}\Delta\Gamma^{p}\right]  _{\lambda=0}=\sigma^{a}%
\Gamma^{p}\ \ \mathrm{and}\mathtt{\ \ }\left[  \frac{\overleftarrow{\partial}%
}{\partial\lambda_{a}}\frac{\overleftarrow{\partial}}{\partial\lambda_{b}%
}\Delta\Gamma^{p}\right]  =\frac{1}{2}\varepsilon^{ab}\sigma^{2}\Gamma
^{p},\ \ \ \mathrm{where}\ \ \ \sigma^{2}=\sigma_{a}\sigma^{a}\ .
\label{Gamma_fin_def}%
\end{equation}
Thus determined finite BRST-antiBRST symmetry transformations for the
integrand $\mathcal{I}_{\Gamma}^{\left(  {F}\right)  }$ in a general gauge
theory have the form ($\mathcal{X}^{pa}=\sigma^{a}\Gamma^{p}\ $and
$\mathcal{Y}^{p}=\left(  1/2\right)  \mathcal{X}_{,q}^{pa}\mathcal{X}%
^{qb}\varepsilon_{ba}=-\left(  1/2\right)  \sigma^{2}\Gamma^{p}$)%
\begin{equation}
\Delta\Gamma^{p}=\mathcal{X}^{pa}\lambda_{a}-\frac{1}{2}\mathcal{Y}^{p}%
\lambda^{2}=\left(  \sigma^{a}\Gamma^{p}\right)  \lambda_{a}+\frac{1}%
{4}\left(  \sigma^{2}\Gamma^{p}\right)  \lambda^{2}\ ,\ \ \ \mathcal{I}%
_{\Gamma+\Delta\Gamma}^{\left(  _{F}\right)  }=\mathcal{I}_{\Gamma}^{\left(
_{F}\right)  }, \label{Gamma_fin}%
\end{equation}
or, in terms of the components,%
\begin{align}
\Delta\phi^{A}  &  =\pi^{Aa}\lambda_{a}+\frac{1}{2}\lambda^{A}\lambda
^{2}%
,\phantom{= \pi^{Aa}\lambda_{a}+ \frac{1}{2}\lambda^{A} \lambda^{2}\quad\quad\ }\Delta
\bar{\phi}_{A}\ =\ \varepsilon^{ab}\lambda_{a}\phi_{Ab}^{\ast}+\frac{1}%
{2}S_{,A}\lambda^{2},\nonumber\\
\Delta\pi^{Aa}  &  =-\varepsilon^{ab}\lambda^{A}\lambda_{b}%
\ ,\phantom{= \pi^{Aa}\lambda_{a}+ \frac{1}{2}\lambda^{A} \lambda^{2}=\varepsilon^{ab}\lambda^{A}\lambda_{a}}\Delta
\lambda^{A}\ =\ 0\ ,\label{exply}\\
\Delta\phi_{Aa}^{\ast}  &  =\lambda_{a}S_{,A}+\frac{1}{4}\left(  -1\right)
^{\varepsilon_{A}}\left[  \varepsilon_{ab}\frac{\delta^{2}S}{\delta\phi
^{A}\delta\phi^{B}}\pi^{Bb}+\varepsilon_{ab} \frac{\delta S}{\delta\phi^{B}%
}\frac{\delta^{2}S}{\delta\phi^{A}\delta\phi_{Bb}^{\ast}} \left(  -1\right)
^{\varepsilon_{B}}-\phi_{Ba}^{\ast}\frac{\delta^{2}S}{\delta\phi^{A}\delta
\bar{\phi}_{B}}\left(  -1\right)  ^{\varepsilon_{B}}\right]  \lambda
^{2}\ .\nonumber
\end{align}

\section*{Acknowledgments}

The authors are grateful to the referee for helpful critical remarks. A.R.
thanks D. Bykov, D. Francia and the participants of the International Seminar
``QUARKS 2014'' for useful remarks and discussions. The study was carried out
within the Tomsk State University Competitiveness Improvement Program and was
supported by the RFBR grant under Project No. 12-02-000121,  by the grant
of Leading Scientific Schools of the Russian Federation under Project No. 88.2014.2, and  partially supported by Ministry of science of  Russian Federation
(grant  No. 2014/223).

\appendix

\section*{Appendix}

\section{Group Properties of Finite BRST-antiBRST Transformations}

\label{AppC} \renewcommand{\theequation}{\Alph{section}.\arabic{equation}} \setcounter{equation}{0}

In this Appendix, in order to clarify the relations (\ref{D1D2F}%
)--(\ref{comm}) of Section~\ref{fBRSTa}, we examine the composition of finite
variations $\Delta_{\left(  1\right)  }\Delta_{\left(  2\right)  }$ acting on
an arbitrary functional $F=F\left(  \phi\right)  $, with the variation $\Delta
F$ given by (\ref{DeltaF}),%
\begin{equation}
\Delta F=\left(  s^{a}F\right)  \lambda_{a}+\frac{1}{4}\left(  s^{2}F\right)
\lambda^{2}\ . \label{ADeltaF}%
\end{equation}
Using the readily established Leibnitz-like properties of the generators of
BRST-antiBRST transformations, $s^{a}$ and $s^{2}$, acting on the product of
any functionals $A$, $B$ with definite Grassmann parities,%
\begin{align}
&  s^{a}\left(  AB\right)  =\left(  s^{a}A\right)  B\left(  -1\right)
^{\varepsilon_{B}}+A\left(  s^{a}B\right)  \,\,\,\,\mathrm{and}\,\,\,\,s_{a}%
\left(  AB\right)  =\left(  s_{a}A\right)  B\left(  -1\right)  ^{\varepsilon
_{B}}+A\left(  s_{a}B\right)  \,,\nonumber\\
&  s^{2}\left(  AB\right)  =\left(  s^{2}A\right)  B-2\left(  s_{a}A\right)
\left(  s^{a}B\right)  \left(  -1\right)  ^{\varepsilon_{B}}+A\left(
s^{2}B\right)  \,,\,\,\,\mathrm{for}\,\,\,s^{2}=s_{a}s^{a}\ , \label{s2(AB)}%
\end{align}
and the identities%
\begin{equation}
s^{a}s^{b}=\left(  1/2\right)  \varepsilon^{ab}s^{2}\,\,\,\,\mathrm{and}%
\,\,\,\,s_{a}s^{b}=-s^{b}s_{a}=\left(  1/2\right)  \delta_{a}^{b}%
s^{2}\,\,\,\,\mathrm{and}\,\,\,\,s^{a}s^{b}s^{c}\equiv0\ , \label{sasb}%
\end{equation}
with the notation $UV\equiv U_{a}V^{a}=-U^{a}V_{a}$ for pairing up any
$\mathrm{Sp}(2)$-vectors $U^{a}$, $V^{a}$, we obtain%
\begin{align}
s^{a}\left(  \Delta F\right)   &  =s^{a}\left[  \left(  s^{b}F\right)
\lambda_{b}+\frac{1}{4}\left(  s^{2}F\right)  \lambda^{2}\right]
=s^{a}\left[  \left(  s^{b}F\right)  \lambda_{b}\right]  +\left(  1/4\right)
s^{a}\left[  \left(  s^{2}F\right)  \lambda^{2}\right] \nonumber\\
&  =-\left(  s^{a}s^{b}F\right)  \lambda_{b}+\left(  s^{b}F\right)  \left(
s^{a}\lambda_{b}\right)  +\left(  1/4\right)  \left(  s^{2}F\right)  \left(
s^{a}\lambda^{2}\right) \nonumber\\
&  =-\left(  1/2\right)  \left(  s^{2}F\right)  \lambda^{a}-\left(  sF\right)
\left(  s^{a}\lambda\right)  +\left(  1/4\right)  \left(  s^{2}F\right)
\left(  s^{a}\lambda^{2}\right)  \label{saDeltaF}%
\end{align}
and%
\begin{align}
s^{2}\left(  \Delta F\right)   &  =s^{2}\left[  \left(  s^{b}F\right)
\lambda_{b}+\frac{1}{4}\left(  s^{2}F\right)  \lambda^{2}\right]
=s^{2}\left[  \left(  s^{b}F\right)  \lambda_{b}\right]  +\frac{1}{4}%
s^{2}\left[  \left(  s^{2}F\right)  \lambda^{2}\right] \nonumber\\
&  =2\left(  s_{a}s^{b}F\right)  \left(  s^{a}\lambda_{b}\right)  +\left(
s^{b}F\right)  \left(  s^{2}\lambda_{b}\right)  +\frac{1}{4}\left(
s^{2}F\right)  \left(  s^{2}\lambda^{2}\right) \nonumber\\
&  =-\left(  s^{2}F\right)  \left(  s\lambda\right)  -\left(  sF\right)
\left(  s^{2}\lambda\right)  +\frac{1}{4}\left(  s^{2}F\right)  \left(
s^{2}\lambda^{2}\right)  \ . \label{s2DeltaF}%
\end{align}
Therefore, $\Delta_{\left(  1\right)  }\Delta_{\left(  2\right)  }F$ is given
by%
\begin{align}
\Delta_{\left(  1\right)  }\Delta_{\left(  2\right)  }F  &  =\left(
s^{a}\Delta_{\left(  2\right)  }F\right)  \lambda_{\left(  1\right)  a}%
+\frac{1}{4}\left(  s^{2}\Delta_{\left(  2\right)  }F\right)  \lambda_{\left(
1\right)  }^{2}\nonumber\\
&  =\left[  -\left(  1/2\right)  \left(  s^{2}F\right)  \lambda_{\left(
2\right)  }^{a}-\left(  sF\right)  \left(  s^{a}\lambda_{\left(  2\right)
}\right)  +\left(  1/4\right)  \left(  s^{2}F\right)  \left(  s^{a}%
\lambda_{\left(  2\right)  }^{2}\right)  \right]  \lambda_{\left(  1\right)
a}\nonumber\\
&  +\frac{1}{4}\left[  \left(  s^{2}F\right)  \left(  s\lambda_{\left(
2\right)  }\right)  -\left(  sF\right)  \left(  s^{2}\lambda_{\left(
2\right)  }\right)  +\frac{1}{4}\left(  s^{2}F\right)  \left(  s^{2}%
\lambda_{\left(  2\right)  }^{2}\right)  \right]  \lambda_{\left(  1\right)
}^{2}\nonumber\\
&  \equiv\left(  s^{a}F\right)  \vartheta_{\left(  1,2\right)  a}+\frac{1}%
{4}\left(  s^{2}F\right)  \theta_{\left(  1,2\right)  }\ , \label{DDeltaF}%
\end{align}
whence%
\begin{align}
\vartheta_{\left(  1,2\right)  }^{a}  &  =-\left(  s\lambda_{\left(  2\right)
}^{a}\right)  \lambda_{\left(  1\right)  }+\frac{1}{4}\left(  s^{2}%
\lambda_{\left(  2\right)  }^{a}\right)  \lambda_{\left(  1\right)  }%
^{2}\ ,\label{vartheta12}\\
\theta_{\left(  1,2\right)  }  &  =\left[  2\lambda_{\left(  2\right)
}-\left(  s\lambda_{\left(  2\right)  }^{2}\right)  \right]  \lambda_{\left(
1\right)  }-\left[  \left(  s\lambda_{\left(  2\right)  }\right)  -\frac{1}%
{4}\left(  s^{2}\lambda_{\left(  2\right)  }^{2}\right)  \right]
\lambda_{\left(  1\right)  }^{2}\ . \label{theta12}%
\end{align}
Hence, the commutator of finite variations reads%
\begin{equation}
\left[  \Delta_{\left(  1\right)  },\Delta_{\left(  2\right)  }\right]
F=\left(  s^{a}F\right)  \vartheta_{\left[  1,2\right]  a}+\frac{1}{4}\left(
s^{2}F\right)  \theta_{\left[  1,2\right]  }\ . \label{[d1d2]F}%
\end{equation}
Finally, using the identity%
\begin{equation}
\lambda_{\left(  2\right)  }\lambda_{\left(  1\right)  }-\lambda_{\left(
1\right)  }\lambda_{\left(  2\right)  }=\lambda_{\left(  2\right)  a}%
\lambda_{\left(  1\right)  }^{a}-\lambda_{\left(  1\right)  a}\lambda_{\left(
2\right)  }^{a}=\lambda_{\left(  2\right)  a}\lambda_{\left(  1\right)  }%
^{a}-\lambda_{\left(  2\right)  a}\lambda_{\left(  1\right)  }^{a}\equiv0\ ,
\label{l1l2}%
\end{equation}
we obtain%
\begin{align}
\vartheta_{\left[  1,2\right]  }^{a}=\vartheta_{\left(  1,2\right)  }%
^{a}-\vartheta_{\left(  2,1\right)  }^{a}=  &  \left(  s\lambda_{\left(
1\right)  }^{a}\right)  \lambda_{\left(  2\right)  }-\left(  s\lambda_{\left(
2\right)  }^{a}\right)  \lambda_{\left(  1\right)  }-\frac{1}{4}\left[
\left(  s^{2}\lambda_{\left(  1\right)  }^{a}\right)  \lambda_{\left(
2\right)  }^{2}-\left(  s^{2}\lambda_{\left(  2\right)  }^{a}\right)
\lambda_{\left(  1\right)  }^{2}\right]  \ ,\label{vartheta[12]}\\
\theta_{\left[  1,2\right]  }=\theta_{\left(  1,2\right)  }-\theta_{\left(
2,1\right)  }=  &  \left[  \left(  s\lambda_{\left(  1\right)  }^{2}\right)
\lambda_{\left(  2\right)  }-\left(  s\lambda_{\left(  2\right)  }^{2}\right)
\lambda_{\left(  1\right)  }\right]  +\left[  \left(  s\lambda_{\left(
1\right)  }\right)  \lambda_{\left(  2\right)  }^{2}-\left(  s\lambda_{\left(
2\right)  }\right)  \lambda_{\left(  1\right)  }^{2}\right] \nonumber\\
&  +\frac{1}{4}\left[  \left(  s^{2}\lambda_{\left(  2\right)  }^{2}\right)
\lambda_{\left(  1\right)  }^{2}-\left(  s^{2}\lambda_{\left(  1\right)  }%
^{2}\right)  \lambda_{\left(  2\right)  }^{2}\right]  \ . \label{theta[12]}%
\end{align}
In particular, the linear approximation $\Delta^{\mathrm{lin}}F=\left(
s^{a}F\right)  \lambda_{a}$,$\ \Delta F=\Delta^{\mathrm{lin}}F+O\left(
\lambda^{2}\right)  $, implies (\ref{comm}).

\section{Calculation of Jacobians}

\label{AppA} \renewcommand{\theequation}{\Alph{section}.\arabic{equation}} \setcounter{equation}{0}

In this Appendix, we present the calculation of the Jacobian (\ref{measure}),
(\ref{superJ}), induced in the functional integral\ (\ref{z(j)}) by the finite
BRST-antiBRST transformations (\ref{finite}) with an $\mathrm{Sp}\left(
2\right)  $-doublet of anticommuting parameters $\lambda_{a}$, considering the
global case, $\lambda_{a}=\mathrm{const}$, and the case of field-dependent
functionals $\lambda_{a}\left(  \phi\right)  $ of a special form, $\lambda
_{a}\left(  \phi\right)  =s_{a}\Lambda\left(  \phi\right)  $.

\subsection{Constant Parameters}

\label{AppA1}

Let us assume $\lambda_{a}$ to be constant parameters in (\ref{finite}) and
consider an even matrix $M$ in (\ref{superJ}) with the elements $M_{B}^{A}$,
$\varepsilon\left(  M_{B}^{A}\right)  =\varepsilon_{A}+\varepsilon_{B}$,%
\begin{equation}
M_{B}^{A}\ =\ \frac{\delta\left(  \Delta\phi^{A}\right)  }{\delta\phi^{B}%
}\ =\ \left(  Q_{1}\right)  _{B}^{A}+R_{B}^{A}\,,\,\,\,\mathrm{with}%
\,\,\,\left(  Q_{1}\right)  _{B}^{A}\ =\ \frac{\delta X^{Aa}}{\delta\phi^{B}%
}\lambda_{a}\left(  -1\right)  ^{\varepsilon_{B}}\,\,\,\mathrm{and}%
\,\,\,R_{B}^{A}=-\frac{1}{2}\frac{\delta Y^{A}}{\delta\phi^{B}}\lambda^{2}\ .
\label{MAB}%
\end{equation}
Notice the fact that $Q_{1}\sim\lambda_{a}$, $R\sim\lambda^{2}$, which, in
view of the nilpotency properties $\lambda_{a}\lambda^{2}=\lambda^{4}\equiv0$,
implies%
\begin{equation}
\mathrm{Str}\left(  M^{n}\right)  =\mathrm{Str}\left(  Q_{1}+R\right)
^{n}=\left\{
\begin{array}
[c]{ll}%
\mathrm{Str}\left(  Q_{1}+R\right)  =\mathrm{Str}\left(  R\right)  \ , &
n=1\ ,\\
\mathrm{Str}\left(  Q_{1}^{2}\right)  =2\mathrm{Str}\left(  R\right)  \ , &
n=2\ ,\\
0\ , & n>2\ .
\end{array}
\right.  \label{repsMAB}%
\end{equation}
Indeed, due to the relations $X_{,A}^{Aa}=0$ in (\ref{xy2}), we have%
\begin{equation}
\mathrm{Str}\left(  Q_{1}\right)  =\left(  Q_{1}\right)  _{A}^{A}\left(
-1\right)  ^{\varepsilon_{A}}=\frac{\delta X^{Aa}}{\delta\phi^{A}}\lambda
_{a}=0\ . \label{Q1}%
\end{equation}
Next, let us examine $\mathrm{Str}\left(  Q_{1}^{2}\right)  $:%
\begin{equation}
\mathrm{Str}\left(  Q_{1}^{2}\right)  =\left(  Q_{1}^{2}\right)  _{A}%
^{A}\left(  -1\right)  ^{\varepsilon_{A}}=\frac{\delta X^{Aa}}{\delta\phi^{B}%
}\lambda_{a}\frac{\delta X^{Bb}}{\delta\phi^{A}}\lambda_{b}\left(  -1\right)
^{\varepsilon_{B}}=\frac{\delta X^{Aa}}{\delta\phi^{B}}\frac{\delta X^{Bb}%
}{\delta\phi^{A}}\lambda_{b}\lambda_{a}\left(  -1\right)  ^{\varepsilon_{A}%
}\ . \label{Q12rel}%
\end{equation}
Differentiating the relation $X_{,B}^{Aa}X^{Bb}=\varepsilon^{ab}Y^{A}$ in
(\ref{xy2})\ with respect to $\phi^{A}$, we find%
\[
\frac{\delta}{\delta\phi^{B}}\left(  \frac{\delta X^{Aa}}{\delta\phi^{A}%
}\right)  X^{Bb}\left(  -1\right)  ^{\varepsilon_{B}}+\frac{\delta X^{Aa}%
}{\delta\phi^{B}}\frac{\delta X^{Bb}}{\delta\phi^{A}}+\varepsilon^{ba}%
\frac{\delta Y^{A}}{\delta\phi^{A}}=0\ .
\]
Then, due to the relation $X_{,A}^{Aa}=0$ in (\ref{xy2}), we have%
\begin{equation}
\frac{\delta X^{Aa}}{\delta\phi^{B}}\frac{\delta X^{Bb}}{\delta\phi^{A}%
}=\varepsilon^{ab}\frac{\delta Y^{A}}{\delta\phi^{A}}\ , \label{conseq}%
\end{equation}
and therefore%
\begin{equation}
\mathrm{Str}\left(  Q_{1}^{2}\right)  =\varepsilon^{ab}\frac{\delta Y^{A}%
}{\delta\phi^{A}}\lambda_{b}\lambda_{a}\left(  -1\right)  ^{\varepsilon_{A}%
}=-\frac{\delta Y^{A}}{\delta\phi^{A}}\lambda^{2}\left(  -1\right)
^{\varepsilon_{A}}=2\mathrm{Str}\left(  R\right)  \ . \label{2R}%
\end{equation}
Thus, the Jacobian $\exp\left(  \Im\right)  $ in (\ref{superJ}) is given by%
\begin{equation}
\Im=-\sum_{n=1}^{\infty}\frac{\left(  -1\right)  ^{n}}{n}\mathrm{Str}\left(
M^{n}\right)  =\mathrm{Str}\left(  M\right)  -\frac{1}{2}\mathrm{Str}\left(
M^{2}\right)  =\mathrm{Str}\left(  R\right)  -\frac{1}{2}\mathrm{Str}\left(
Q_{1}^{2}\right)  \equiv0\ , \label{zero}%
\end{equation}
which proves (\ref{constJ}).

\subsection{Field-dependent Parameters}

\label{AppA2}

In the case of field-dependent parameters $\lambda_{a}\left(  \phi\right)
=s_{a}\Lambda\left(  \phi\right)  $ from (\ref{finite}), given by an
even-valued potential $\Lambda\left(  \phi\right)  $, let us consider an even
matrix $M$ in (\ref{superJ}) with the elements $M_{B}^{A}$,%
\begin{align}
&  M_{B}^{A}\equiv\frac{\delta\left(  \Delta\phi^{A}\right)  }{\delta\phi^{B}%
}=P_{B}^{A}+Q_{B}^{A}+R_{B}^{A}\,,\,\,\,\mathrm{with}\,\,\,Q_{B}^{A}=\left(
Q_{1}\right)  _{B}^{A}+\left(  Q_{2}\right)  _{B}^{A}\ ,\label{MABext}\\
&  \mathrm{for}\,\,\,P_{B}^{A}=X^{Aa}\frac{\delta\lambda_{a}}{\delta\phi^{B}%
}\ ,\quad\left(  Q_{1}\right)  _{B}^{A}=\lambda_{a}\frac{\delta X^{Aa}}%
{\delta\phi^{B}}\left(  -1\right)  ^{\varepsilon_{A}+1}\ ,\quad\left(
Q_{2}\right)  _{B}^{A}=\lambda_{a}Y^{A}\frac{\delta\lambda^{a}}{\delta\phi
^{B}}\left(  -1\right)  ^{\varepsilon_{A}+1}\ ,\quad R_{B}^{A}=-\frac{1}%
{2}\lambda^{2}\frac{\delta Y^{A}}{\delta\phi^{B}}. \label{PRQABext}%
\end{align}
Using the property
\begin{equation}
\mathrm{Str}\left(  AB\right)  =\mathrm{Str}\left(  BA\right)  \,,
\label{transp}%
\end{equation}
which takes place for any even matrices $A,B$, and the fact that the
occurrence of $R\sim\lambda^{2}$ in $\mathrm{Str}\left(  M^{n}\right)  $ more
than once yields zero, $\lambda^{4}\equiv0$, we have%
\begin{equation}
\mathrm{Str}\left(  M^{n}\right)  =\mathrm{Str}\left(  P+Q+R\right)  ^{n}%
=\sum_{k=0}^{1}C_{n}^{k}\mathrm{Str}\left[  \left(  P+Q\right)  ^{n-k}%
R^{k}\right]  \ ,\ \ \ C_{n}^{k}=\frac{n!}{k!\left(  n-k\right)  !}\ .
\label{Strgen}%
\end{equation}
Furthermore,%
\begin{equation}
\mathrm{Str}\left(  P+Q+R\right)  ^{n}=\mathrm{Str}\left(  P+Q\right)
^{n}+n\mathrm{Str}\left[  \left(  P+Q\right)  ^{n-1}R\right]  =\mathrm{Str}%
\left(  P+Q\right)  ^{n}+n\mathrm{Str}\left(  P^{n-1}R\right)  \ , \label{R}%
\end{equation}
since any occurrence of $R\sim\lambda^{2}$ and $Q\sim\lambda_{a}$
simultaneously entering $\mathrm{Str}\left(  M\right)  ^{n}$\ yields zero,
owing to $\lambda_{a}\lambda^{2}=0$, as a consequence of which $R$ can only be
coupled with $P^{n-1}$.

Having established (\ref{R}), let us examine $\mathrm{Str}\left(
P^{n-1}R\right)  $, namely,%
\begin{equation}
\mathrm{Str}\left(  P^{n-1}R\right)  =\left\{
\begin{array}
[c]{ll}%
\mathrm{Str}\left(  R\right)  \ , & n=1\,,
\\
0\ , & n>1\,.
\end{array}
\right.  \label{StrPR}%
\end{equation}
Indeed, due to the contraction property $P^{2}=f\cdot P\Longrightarrow
P^{l}=f^{l-1}\cdot P$, where $f$ is an even-valued parameter (for details, see
(\ref{contracprop}) below), we have
\begin{align}
&  \mathrm{Str}\left(  P^{n-1}R\right)  =f^{n-2}\mathrm{Str}\left(  PR\right)
\ ,\ \ n>1\ ,\label{StrPR1}\\
&  \mathrm{Str}\left(  PR\right)  =\mathrm{Str}\left(  RP\right)  =\left(
RP\right)  _{A}^{A}\left(  -1\right)  ^{\varepsilon_{A}}=R_{B}^{A}P_{A}%
^{B}\left(  -1\right)  ^{\varepsilon_{A}}=-\frac{1}{2}\lambda^{2}\left(
\frac{\delta Y^{A}}{\delta\phi^{B}}X^{Bb}\right)  \frac{\delta\lambda_{b}%
}{\delta\phi^{A}}\left(  -1\right)  ^{\varepsilon_{A}}=0\ , \label{StrPR2}%
\end{align}
since $Y_{,B}^{A}X^{Bb}=0$ in (\ref{xy2}), which implies%
\begin{equation}
\mathrm{Str}\left(  M^{n}\right)  =\mathrm{Str}\left(  P+Q\right)
^{n}+n\mathrm{Str}\left(  P^{n-1}R\right)  =\left\{
\begin{array}
[c]{ll}%
\mathrm{Str}\left(  P+Q\right)  +\mathrm{Str}\left(  R\right)  \ , & n=1\ ,\\
\mathrm{Str}\left(  P+Q\right)  ^{n}\ , & n>1\ ,
\end{array}
\right.  \label{P+Q}%
\end{equation}
so that $R$ drops out of $\mathrm{Str}\left(  M^{n}\right)  $, $n>1,$ and
enters the Jacobian only as $\mathrm{Str}\left(  R\right)  $.

Considering the contribution $\mathrm{Str}\left(  P+Q\right)  ^{n}$ in
(\ref{P+Q}), we notice that an occurrence of $Q\sim\lambda_{a}$ more then
twice yields zero, $\lambda_{a}\lambda_{b}\lambda_{c}\equiv0$. A direct
calculation for $n=2,3$ leads to%
\begin{equation}
\mathrm{Str}\left(  P+Q\right)  ^{n}=\sum_{k=0}^{n}C_{n}^{k}\mathrm{Str}%
\left(  P^{n-k}Q^{k}\right)  =\mathrm{Str}\left(  P^{n}+nP^{n-1}Q+C_{n}%
^{2}P^{n-2}Q^{2}\right)  \ . \label{StrP+Q}%
\end{equation}
Next, starting from the case $n=4$, $\mathrm{Str}\left(  M^{4}\right)
=\mathrm{Str}\left(  P^{4}+4P^{3}Q+4P^{2}Q^{2}+2PQPQ\right)  $, one can prove
that for any $n\geq4\ $we have
\begin{equation}
\mathrm{Str}\left(  P+Q\right)  ^{n}=\mathrm{Str}\left(  P^{n}+nP^{n-1}%
Q+nP^{n-2}Q^{2}+K_{n}P^{n-3}QPQ\right)  \ , \label{Str(P+Q)^n}%
\end{equation}
where the coefficients\footnote{The coefficient $K_{n}$ turns out to be the
number of monomials in $\left(  P+Q\right)  ^{n}$\ for $n\geq4$ that contain
two matrices $Q$\ and cannot be transformed by cyclic permutations under the
symbol $\mathrm{Str}$ of supertrace to the form $\mathrm{Str}(P^{n-2}Q^{2})$.}
$K_{n}$ are given by (in particular, $n=4$,$\ C_{4}^{2}=6$,$\ K_{4}=C_{4}%
^{2}-4=2$)%
\begin{equation}
K_{n}=C_{n}^{2}-n\ ,\ \ \ C_{n}^{2}=n\left(  n-1\right)  /2\ \Longrightarrow
\ K_{n}=n\left(  n-3\right)  /2\ , \label{Kn}%
\end{equation}
which implies%
\begin{equation}
\frac{C_{n}^{2}}{n}-\frac{K_{n}}{n}=1\ ,\ \ \ \frac{C_{n}^{2}}{n}%
-\frac{K_{n+1}}{n+1}=\frac{1}{2}\ . \label{C,K}%
\end{equation}
The proof of (\ref{Str(P+Q)^n}) goes by induction. To this end, suppose that
(as in the case $n=4$)%
\begin{align}
&  \left(  P+Q\right)  ^{n}=P^{n}+A_{n}^{\left(  1\right)  }\left(
P,Q\right)  +B_{n}^{\left(  2\right)  }\left(  P,Q\right)  +C_{n}^{\left(
2\right)  }\left(  P,Q\right)  \,,\,\,\,\mathrm{where}\nonumber\\
&  A_{n}^{\left(  1\right)  }=a_{kl}P^{k}QP^{l}\ ,\ \ \ a_{n}\equiv
a_{k0}=1\ ,\ \ \ B_{n}^{\left(  2\right)  }=b_{kl}P^{k}Q^{2}P^{l}%
\ ,\ \ \ C_{n}^{\left(  2\right)  }=c_{kml}P^{k}QP^{m}QP^{l}\ ,\ \ \ m\geq
1\ ,\nonumber\\
&  \mathrm{and}\,\,\,\mathrm{Str}\left(  A_{n}^{\left(  1\right)  }\right)
=n\mathrm{Str}\left(  P^{n-1}Q\right)  \ ,\ \ \ \mathrm{Str}\left(
B_{n}^{\left(  2\right)  }\right)  =n\mathrm{Str}\left(  P^{n-2}Q^{2}\right)
\ ,\ \ \ \mathrm{Str}\left(  C_{n}^{\left(  2\right)  }\right)  =K_{n}%
\mathrm{Str}\left(  P^{n-3}QPQ\right)  \ . \label{suppose}%
\end{align}
Then, due to the vanishing of the terms containing $Q$ more than twice, we
have%
\begin{align}
&  \left(  P+Q\right)  ^{n+1}=P^{n+1}+A_{n+1}^{\left(  1\right)  }%
+B_{n+1}^{\left(  2\right)  }+C_{n+1}^{\left(  2\right)  }\ ,\nonumber\\
&  \mathrm{for}\,\,\,A_{n+1}^{\left(  1\right)  }=P^{n}Q+A_{n}^{\left(
1\right)  }P\ ,\ \ \ B_{n+1}^{\left(  2\right)  }+C_{n+1}^{\left(  2\right)
}=A_{n}^{\left(  1\right)  }Q+B_{n}^{\left(  2\right)  }P+C_{n}^{\left(
2\right)  }P\ , \label{ind_conseq}%
\end{align}
where%
\begin{align}
A_{n+1}^{\left(  1\right)  }  &  =P^{n}Q+a_{kl}P^{k}QP^{l}P\ \Longrightarrow
\ a_{n+1}=1\ ,\label{A(n+1)}\\
B_{n+1}^{\left(  2\right)  }  &  =a_{k0}P^{k}Q^{2}+B_{n}^{\left(  2\right)
}P\ ,\ \ \ C_{n+1}^{\left(  2\right)  }=a_{kl}P^{k}QP^{l}Q+C_{n}^{\left(
2\right)  }P\ ,\ \ \ l\geq1\ . \label{B,C(n+1)}%
\end{align}
Due to the contraction property $P^{2}=f\cdot P\Longrightarrow P^{l}%
=f^{l-1}\cdot P$ in (\ref{contracprop}), the above implies%
\begin{align}
\mathrm{Str}\left(  A_{n+1}^{\left(  1\right)  }\right)   &  =\left(
n+1\right)  \mathrm{Str}\left(  P^{n}Q\right)  \ ,\ \ \ \mathrm{Str}\left(
B_{n+1}^{\left(  2\right)  }\right)  =\left(  n+1\right)  \mathrm{Str}\left(
P^{n}Q^{2}\right)  \ ,\label{strA,B(n+1)}\\
\mathrm{Str}\left(  C_{n+1}^{\left(  2\right)  }\right)   &  =\left(
n-1\right)  \mathrm{Str}\left(  P^{n-2}QPQ\right)  +K_{n}\mathrm{Str}\left(
P^{n-2}QPQ\right)  \ . \label{strC(n+1)}%
\end{align}
Notice that%
\begin{equation}
K_{n}+n-1=\frac{n\left(  n-3\right)  }{2}+\frac{2n-2}{2}=\frac{\left(
n+1\right)  \left(  n-2\right)  }{2}=K_{n+1}\ , \label{propKn}%
\end{equation}
which proves the induction.

Recall that the Jacobian $\exp\left(  \Im\right)  $ in (\ref{superJ}) is given
by
\begin{equation}
\Im=\mathrm{Str}\ln\left(  \mathbb{I}+M\right)  =-\sum_{n=1}^{\infty}%
\frac{\left(  -1\right)  ^{n}}{n}\,\,\mathrm{Str}\left(  M^{n}\right)  \,,
\label{superJagain}%
\end{equation}
where, according to the previous considerations,%
\begin{align}
&  \mathrm{Str}\left(  M^{n}\right)  =\sum_{k=0}^{1}C_{n}^{k}\mathrm{Str}%
\left(  P^{n-k}Q^{k}\right)  +D_{n}\ ,\ \ \ n\geq1\ ,\label{M^n}\\
&  \mathrm{for}\,\,\,D_{n}=\left\{
\begin{array}
[c]{ll}%
\mathrm{Str}\left(  R\right) \ , & n=1\ ,\\
C_{n}^{2}\mathrm{Str}\left(  P^{n-2}Q^{2}\right)  \ , & n=2,3\ ,\\
\left(  C_{n}^{2}-K_{n}\right)  \mathrm{Str}\left(  P^{n-2}Q^{2}\right)
+K_{n}\mathrm{Str}\left(  P^{n-3}QPQ\right)  \ , & n>3\ ,
\end{array}
\right.  \label{D_n}%
\end{align}
or, in detail,%
\begin{equation}
\mathrm{Str}\left(  M^{n}\right)  =\left\{
\begin{array}
[c]{ll}%
\mathrm{Str}\left(  P\right)  +\mathrm{Str}\left(  Q\right)  +\mathrm{Str}%
\left(  R\right)  \ , & n=1\ ,\\
\mathrm{Str}\left(  P^{n}\right)  +C_{n}^{1}\mathrm{Str}\left(  P^{n-1}%
Q\right)  +C_{n}^{2}\mathrm{Str}\left(  P^{n-2}Q^{2}\right)  \ , & n=2,3\ ,\\
\mathrm{Str}\left(  P^{n}\right)  +C_{n}^{1}\mathrm{Str}\left(  P^{n-1}%
Q\right)  +\left(  C_{n}^{2}-K_{n}\right)  \mathrm{Str}\left(  P^{n-2}%
Q^{2}\right)  +K_{n}\mathrm{Str}\left(  P^{n-3}QPQ\right)  \ , & n>3\ .
\end{array}
\right.  \label{str(M^n)}%
\end{equation}

First of all, the calculation of the Jacobian is based on the previously
established properties (\ref{2R}) and (\ref{Q1}), namely,%
\begin{equation}
\mathrm{Str}\left(  Q_{1}\right)  =0\ ,\ \ \ \mathrm{Str}\left(  Q_{1}%
^{2}\right)  =2\mathrm{Str}\left(  R\right)  \ . \label{prev_prop}%
\end{equation}
It has also been established (Appendix \ref{AppA1}) that the quantity
$\mathrm{Str}\left(  R\right)  $ in (\ref{P+Q}) cancels the contribution
$\mathrm{Str}\left(  Q_{1}^{2}\right)  $ to the Jacobian, where these
contributions enter in the first and second orders, $\mathrm{Str}\left(
M^{1}\right)  $ and $\mathrm{Str}\left(  M^{2}\right)  $, respectively, thus
summarily producing an identical zero:%
\begin{equation}
\mathrm{Str}\left(  R\right)  -\left(  1/2\right)  \mathrm{Str}\left(
Q_{1}^{2}\right)  \equiv0\ . \label{Q12R}%
\end{equation}
Therefore, we can exclude $\mathrm{Str}\left(  R\right)  $ and $\mathrm{Str}%
\left(  Q_{1}^{2}\right)  $ from further consideration.

Recalling that $\lambda_{a}=s_{a}\Lambda$, we can deduce the additional
properties%
\begin{equation}
P^{2}=f\cdot P\ ,\ \ \ QP=\left(  1+f\right)  \cdot Q_{2}\ ,\ \ \ \ f=-\frac
{1}{2}\mathrm{Str}\left(  P\right)  \ , \label{contracprop}%
\end{equation}
where the quantity $f$ is given by%
\begin{equation}
\frac{\delta\lambda_{b}}{\delta\phi^{A}}X^{Aa}=s^{a}\lambda_{b}=\delta_{b}%
^{a}f\ \Longrightarrow\ f=\frac{1}{2}s^{a}\lambda_{a}=-\frac{1}{2}s^{2}%
\Lambda\ \ . \label{f}%
\end{equation}
Indeed,%
\begin{align}
&  \left(  P^{2}\right)  _{B}^{A}=\left(  P\right)  _{D}^{A}\left(  P\right)
_{B}^{D}=X^{Aa}\left(  \frac{\delta\lambda_{a}}{\delta\phi^{D}}X^{Db}\right)
\frac{\delta\lambda_{b}}{\delta\phi^{B}}=f\cdot\delta_{a}^{b}X^{Aa}%
\frac{\delta\lambda_{b}}{\delta\phi^{B}}=f\cdot\left(  P\right)  _{B}%
^{A}\ ,\nonumber\\
&  \frac{\delta\lambda_{a}}{\delta\phi^{B}}X^{Bb}=s^{b}\lambda_{a}=s^{b}%
s_{a}\Lambda=\delta_{a}^{b}f\ ,\ \quad\ f=\Lambda_{,A}Y^{A}-\left(
1/2\right)  \varepsilon_{ab}X^{Aa}\Lambda_{,AB}X^{Bb}\ ,\nonumber\\
&  f=\frac{1}{2}\left(  \frac{\delta\lambda_{a}}{\delta\phi^{A}}X^{Aa}\right)
=-\frac{1}{2}\left(  P\right)  _{A}^{A}\left(  -1\right)  ^{\varepsilon_{A}%
}=-\frac{1}{2}\mathrm{Str}\left(  P\right)  \ . \label{fP2}%
\end{align}
As a consequence, we have $QP=\left(  1+f\right)  \cdot Q_{2}$, namely, in
view of $X_{,B}^{Aa}X^{Bb}=\varepsilon^{ab}Y^{A}$ from (\ref{xy2}),
\begin{align}
\left(  QP\right)  _{B}^{A}  &  =Q_{D}^{A}P_{B}^{D}=\left(  -1\right)
^{\varepsilon_{A}+1}\lambda_{a}\left(  \frac{\delta X^{Aa}}{\delta\phi^{D}%
}+Y^{A}\frac{\delta\lambda^{a}}{\delta\phi^{D}}\right)  X^{Dd}\frac
{\delta\lambda_{d}}{\delta\phi^{B}}\nonumber\\
&  =\left(  -1\right)  ^{\varepsilon_{A}+1}\lambda_{a}\left[  \varepsilon
^{ab}Y^{A}+Y^{A}\left(  s^{b}\lambda^{a}\right)  \right]  \frac{\delta
\lambda_{b}}{\delta\phi^{B}}=\left(  -1\right)  ^{\varepsilon_{A}+1}%
\lambda_{a}\left[  \varepsilon^{ab}Y^{A}+\varepsilon^{ad}Y^{A}\delta_{d}%
^{b}f\right]  \frac{\delta\lambda_{b}}{\delta\phi^{B}}\nonumber\\
&  =\left(  -1\right)  ^{\varepsilon_{A}+1}\lambda_{a}Y^{A}\left(  1+f\right)
\frac{\delta\lambda^{a}}{\delta\phi^{B}}=\left(  1+f\right)  \left(
Q_{2}\right)  _{B}^{A}\ . \label{fQ2}%
\end{align}
Finally,%
\begin{equation}%
\begin{array}
[c]{lc}%
\mathrm{Str}\left(  P^{n}\right)  =f^{n-1}\mathrm{Str}\left(  P\right)
=-2f^{n}\ , & n\geq1\ ,\\
\mathrm{Str}\left(  P^{n-1}Q\right)  =\left\{
\begin{array}
[c]{l}%
\mathrm{Str}\left(  Q\right)  =\mathrm{Str}\left(  Q_{2}\right)  \ ,\\
f^{n-2}\mathrm{Str}\left(  PQ\right)  =f^{n-2}\mathrm{Str}\left(  QP\right)
=f^{n-2}\left(  1+f\right)  \mathrm{Str}\left(  Q_{2}\right)  \ ,
\end{array}
\right.  &
\begin{array}
[c]{l}%
n=1\ ,\\
n>1\ ,
\end{array}
\\
\mathrm{Str}\left(  P^{n-2}Q^{2}\right)  =\left\{
\begin{array}
[c]{l}%
\mathrm{Str}\left(  Q^{2}\right)  =\mathrm{Str}\left(  2Q_{1}Q_{2}+Q_{2}%
^{2}\right)  \ ,\\
f^{n-3}\mathrm{Str}\left(  PQ^{2}\right)  =f^{n-3}\mathrm{Str}\left[  Q\left(
QP\right)  \right]  =f^{n-3}\left(  1+f\right)  \mathrm{Str}\left[  \left(
Q_{1}+Q_{2}\right)  Q_{2}\right]  \ ,
\end{array}
\right.  &
\begin{array}
[c]{l}%
n=2\ ,\\
n>2\ ,
\end{array}
\\
\mathrm{Str}\left(  P^{n-3}QPQ\right)  =f^{n-4}\mathrm{Str}\left(
PQPQ\right)  =f^{n-4}\mathrm{Str}\left[  \left(  QP\right)  \left(  QP\right)
\right]  =f^{n-4}\left(  1+f\right)  ^{2}\mathrm{Str}\left(  Q_{2}^{2}\right)
\ , & n>3\ ,
\end{array}
\label{strPnQk}%
\end{equation}
where the term $\mathrm{Str}\left(  Q_{1}^{2}\right)  $ has been omitted
according to the previous considerations related to (\ref{Q12R}).

We further notice that $\mathrm{Str}\left(  Q_{1}Q_{2}\right)  \not \equiv 0$.
Indeed, due to $X_{,B}^{Aa}X^{Bb}=\varepsilon^{ab}Y^{A}$ and $Y_{,B}^{A}%
X^{Bb}=0$ in (\ref{xy2}), we have$\,$%
\begin{align}
\left(  Q_{1}Q_{2}\right)  _{A}^{A}\left(  -1\right)  ^{\varepsilon_{A}}  &
=\lambda_{a}\frac{\delta X^{Aa}}{\delta\phi^{B}}Y^{B}\frac{\delta\lambda^{2}%
}{\delta\phi^{A}}=\frac{1}{2}\lambda_{a}\left(  \frac{\delta X^{Aa}}%
{\delta\phi^{B}}\frac{\delta X^{Bb}}{\delta\phi^{D}}\right)  X^{Dd}%
\varepsilon_{db}\frac{\delta\lambda^{2}}{\delta\phi^{A}}\nonumber\\
&  =\frac{1}{2}\lambda_{a}\left[  \frac{\delta}{\delta\phi^{D}}\left(
\frac{\delta X^{Aa}}{\delta\phi^{B}}X^{Bb}\right)  -\left(  \frac{\delta
}{\delta\phi^{D}}\frac{\delta X^{Aa}}{\delta\phi^{B}}\right)  X^{Bb}\left(
-1\right)  ^{\varepsilon_{D}\left(  \varepsilon_{B}+1\right)  }\right]
X^{Dd}\varepsilon_{db}\frac{\delta\lambda^{2}}{\delta\phi^{A}}\nonumber\\
&  =\frac{1}{2}\lambda_{a}\left[  \varepsilon^{ab}\frac{\delta Y^{A}}%
{\delta\phi^{D}}X^{Dd}-\left(  \frac{\delta}{\delta\phi^{D}}\frac{\delta
X^{Aa}}{\delta\phi^{B}}\right)  X^{Bb}X^{Dd}\left(  -1\right)  ^{\varepsilon
_{D}\left(  \varepsilon_{B}+1\right)  }\right]  \varepsilon_{db}\frac
{\delta\lambda^{2}}{\delta\phi^{A}}\nonumber\\
&  =\frac{1}{2}\left(  X^{Bb}\frac{\delta^{2}X^{Aa}}{\delta\phi^{D}\delta
\phi^{B}}X^{Dd}\varepsilon_{db}\right)  \lambda_{a}\frac{\delta\lambda^{2}%
}{\delta\phi^{A}}\ . \label{Q1Q2}%
\end{align}
Besides,%
\begin{equation}
\mathrm{Str}\left(  Q_{2}^{2}\right)  =\mathrm{Str}^{2}\left(  Q_{2}\right)
\not \equiv 0\ . \label{strq22}%
\end{equation}
Indeed,%
\begin{align}
&  \ \left(  Q_{2}\right)  _{A}^{A}\left(  -1\right)  ^{\varepsilon_{A}%
}=\lambda_{a}Y^{A}\frac{\delta\lambda^{a}}{\delta\phi^{A}}\ ,\label{indeed1}\\
&  \ \left(  Q_{2}\right)  _{B}^{A}\left(  Q_{2}\right)  _{A}^{B}\left(
-1\right)  ^{\varepsilon_{A}}=\left(  \lambda_{a}Y^{B}\frac{\delta\lambda^{a}%
}{\delta\phi^{B}}\right)  \left(  \lambda_{b}Y^{A}\frac{\delta\lambda^{b}%
}{\delta\phi^{A}}\right)  \ . \label{indeed2}%
\end{align}
Therefore, $\Im$ in the expression (\ref{superJagain}) for the Jacobian
$\exp\left(  \Im\right)  $ has the general structure%
\begin{align}
&  \Im=A\left(  f\right)  +B\left(  f|Q_{2}\right)  +C\left(  f|Q_{1}%
Q_{2}\right)  \ ,\label{strIm}\\
&  \mathrm{for}\,\,\,B\left(  f|Q_{2}\right)  =b_{1}\left(  f\right)
\mathrm{Str}\left(  Q_{2}\right)  +b_{2}\left(  f\right)  \mathrm{Str}\left(
Q_{2}^{2}\right)  =\left[  b_{1}\left(  f\right)  +b_{2}\left(  f\right)
\mathrm{Str}\left(  Q_{2}\right)  \right]  \mathrm{Str}\left(  Q_{2}\right)
\ ,\nonumber\\
&  \mathrm{and}\,\,\,C\left(  f|Q_{1}Q_{2}\right)  =c\left(  f\right)
\mathrm{Str}\left(  Q_{1}Q_{2}\right)  \ .\nonumber
\end{align}

Let us examine $A\left(  f\right)  $, namely,%
\begin{equation}
\label{Af}A\left(  f\right)  =-\sum_{n=1}^{\infty}\frac{\left(  -1\right)
^{n}}{n}\mathrm{Str}\left(  P^{n}\right)  =2\sum_{n=1}^{\infty}\frac{\left(
-1\right)  ^{n}}{n}f^{n}=-2\ln\left(  1+f\right)  \, .
\end{equation}

Let us examine the explicit structure of the series related to $b_{1}\left(
f\right)  $: the quantity $\mathrm{Str}\left(  Q_{2}\right)  $ derives from
$\mathrm{Str}\left(  P^{n-1}Q\right)  $ for $n\geq1$ in (\ref{strPnQk}), and
is coupled with the combinatorial coefficient $C_{n}^{1}$. The part of $\Im$
containing $\mathrm{Str}\left(  Q_{2}\right)  $ is given by%
\begin{equation}
b_{1}\left(  f\right)  \mathrm{Str}\left(  Q_{2}\right)  =C_{1}^{1}%
\mathrm{Str}\left(  Q_{2}\right)  -\sum_{n=2}^{\infty}\frac{\left(  -1\right)
^{n}}{n}C_{n}^{1}f^{n-2}\left(  1+f\right)  \mathrm{Str}\left(  Q_{2}\right)
\ , \label{b1f}%
\end{equation}
whence%
\begin{equation}
b_{1}\left(  f\right)  =1-\left(  1+f\right)  \sum_{m=0}^{\infty}\left(
-1\right)  ^{m}f^{m}=1-\left(  1+f\right)  \left(  1+f\right)  ^{-1}\equiv0\ .
\label{b1ffin}%
\end{equation}

Let us examine the explicit structure of the series related to $b_{2}\left(
f\right)  $: the quantity $\mathrm{Str}^{2}\left(  Q_{2}\right)  $ derives
from $\mathrm{Str}\left(  P^{n-2}Q^{2}\right)  $ for $n\geq2$ in
(\ref{strPnQk}), coupled with the combinatorial coefficients $C_{n}^{2}$ for
$n=2,3$ and $\left(  C_{n}^{2}-K_{n}\right)  $ for $n>3$, and also derives
from $\mathrm{Str}\left(  P^{n-3}QPQ\right)  $ for $n>3$ in (\ref{strPnQk}),
coupled with the combinatorial coefficients $K_{n}$. The part of $\Im$
containing $\mathrm{Str}^{2}\left(  Q_{2}\right)  $ reads%
\begin{align}
b_{2}\left(  f\right)  \mathrm{Str}^{2}\left(  Q_{2}\right)  =  &
-\frac{\left(  -1\right)  ^{2}}{2}C_{2}^{2}\mathrm{Str}^{2}\left(
Q_{2}\right)  -\frac{\left(  -1\right)  ^{3}}{3}C_{3}^{2}\left(  1+f\right)
\mathrm{Str}^{2}\left(  Q_{2}\right) \nonumber\\
&  -\sum_{n=4}^{\infty}\frac{\left(  -1\right)  ^{n}}{n}\left(  C_{n}%
^{2}-K_{n}\right)  f^{n-3}\left(  1+f\right)  \mathrm{Str}^{2}\left(
Q_{2}\right)  -\sum_{n=4}^{\infty}\frac{\left(  -1\right)  ^{n}}{n}%
K_{n}f^{n-4}\left(  1+f\right)  ^{2}\mathrm{Str}^{2}\left(  Q_{2}\right)  \ ,
\label{b2f}%
\end{align}
whence%
\begin{align}
b_{2}\left(  f\right)   &  =-\frac{1}{2}+\left(  1+f\right)  -\sum
_{n=4}^{\infty}\frac{\left(  -1\right)  ^{n}}{n}\left[  \left(  C_{n}%
^{2}-K_{n}\right)  f^{n-3}\left(  1+f\right)  +K_{n}f^{n-4}\left(  1+f\right)
^{2}\right] \nonumber\\
&  =\frac{1}{2}+f-\left(  1+f\right)  \sum_{n=4}^{\infty}\frac{\left(
-1\right)  ^{n}}{n}\left(  C_{n}^{2}f^{n-3}+K_{n}f^{n-4}\right) \nonumber\\
&  =\frac{1}{2}+f-\left(  1+f\right)  \left[  \frac{1}{2}-\sum_{m=1}^{\infty
}\left(  -1\right)  ^{m}\left(  \frac{C_{m+3}^{2}}{m+3}-\frac{K_{m+4}}%
{m+4}\right)  f^{m}\right]  \ . \label{b2ffin}%
\end{align}
By virtue of (\ref{C,K}), this implies the vanishing of $b_{2}\left(
f\right)  $, namely,
\begin{align}
b_{2}\left(  f\right)   &  =\frac{1}{2}f+\left(  1+f\right)  \sum
_{m=1}^{\infty}\left(  -1\right)  ^{m}\left(  \frac{1}{2}\right)  f^{m}%
=\frac{1}{2}f+\frac{1}{2}\left(  1+f\right)  \sum_{m=1}^{\infty}\left(
-1\right)  ^{m}f^{m}\nonumber\\
&  =\frac{1}{2}f+\frac{1}{2}\left(  1+f\right)  \left[  \left(  1+f\right)
^{-1}-1\right]  \equiv0\ . \label{b2ffin0}%
\end{align}

Let us examine the explicit structure of the series related to $c\left(
f\right)  $: the quantity $\mathrm{Str}\left(  Q_{1}Q_{2}\right)  $ derives
from $\mathrm{Str}\left(  P^{n-2}Q^{2}\right)  $ for $n\geq2$ in
(\ref{strPnQk}), and is coupled with the combinatorial coefficients $C_{n}%
^{2}$, for $n=2,3 $, and $C_{n}^{2}-K_{n}$, for $n>3$. The part of $\Im$
containing $\mathrm{Str}\left(  Q_{1}Q_{2}\right)  $ is given by%
\begin{align}
c\left(  f\right)  \mathrm{Str}\left(  Q_{1}Q_{2}\right)  =  &  -\frac{\left(
-1\right)  ^{2}}{2}C_{2}^{2}\mathrm{Str}\left(  2Q_{1}Q_{2}\right)
-\frac{\left(  -1\right)  ^{3}}{3}C_{3}^{2}\left(  1+f\right)  \mathrm{Str}%
\left(  Q_{1}Q_{2}\right) \nonumber\\
&  -\sum_{n=4}^{\infty}\frac{\left(  -1\right)  ^{n}}{n}\left(  C_{n}%
^{2}-K_{n}\right)  f^{n-3}\left(  1+f\right)  \mathrm{Str}\left(  Q_{1}%
Q_{2}\right)  \ , \label{cf_ini}%
\end{align}
whence%
\begin{equation}
c\left(  f\right)  =-1+\left(  1+f\right)  -\sum_{n=4}^{\infty}\frac{\left(
-1\right)  ^{n}}{n}\left(  C_{n}^{2}-K_{n}\right)  f^{n-3}\left(  1+f\right)
=f-\left(  1+f\right)  \sum_{n=4}^{\infty}\left(  -1\right)  ^{n}\left(
\frac{C_{n}^{2}}{n}-\frac{K_{n}}{n}\right)  f^{n-3}\ . \label{cf}%
\end{equation}
By virtue of (\ref{C,K}), this implies the vanishing of $c\left(  f\right)  $,
namely,
\begin{equation}
c\left(  f\right)  =f-\left(  1+f\right)  \sum_{n=4}^{\infty}\left(
-1\right)  ^{n}f^{n-3}=f+\left(  1+f\right)  \sum_{m=1}^{\infty}\left(
-1\right)  ^{m}f^{m}=f+\left(  1+f\right)  \left[  \left(  1+f\right)
^{-1}-1\right]  \equiv0\ . \label{cffin}%
\end{equation}
From the vanishing of all the coefficients $b_{1}\left(  f\right)  $,
$b_{2}\left(  f\right)  $, $c\left(  f\right)  $, due to (\ref{b1ffin}),
(\ref{b2ffin0}), (\ref{cffin}), we conclude that%
\begin{equation}
B\left(  f|Q_{2}\right)  =b_{1}\left(  f\right)  \mathrm{Str}\left(
Q_{2}\right)  +b_{2}\left(  f\right)  \mathrm{Str}\left(  Q_{2}^{2}\right)
\equiv0\,\,\,\,\mathrm{and}\,\,\,C\left(  f|Q_{1}Q_{2}\right)  =c\left(
f\right)  \mathrm{Str}\left(  Q_{1}Q_{2}\right)  \equiv0\ , \label{BCf}%
\end{equation}
and therefore the Jacobian $\exp\left(  \Im\right)  $ is finally given by%
\begin{equation}
\Im=A\left(  f\right)  +B\left(  f|Q_{2}\right)  +C\left(  f|Q_{1}%
Q_{2}\right)  =A\left(  f\right)  =-2\mathrm{\ln}\left(  1+f\right)
\,\,\,\,\mathrm{for}\,\,\,f=-\left(  1/2\right)  s^{2}\Lambda\,,
\label{finsuperJ}%
\end{equation}
which is identical with (\ref{superJaux}).

\section{BRST-antiBRST Invariant Yang--Mills Action in $R_{\xi}$-like Gauges}

\label{AppB} \renewcommand{\theequation}{\Alph{section}.\arabic{equation}} \setcounter{equation}{0}

In this Appendix, we present the details of calculations used in
Section~\ref{YMgauges} to establish a correspondence between the gauge-fixing
procedures in the Yang--Mills theory described by a gauge-fixing function
$\chi(\phi) = 0$ from the class of $R_{\xi}$-gauges in the BV formalism
\cite{BV} and by a gauge-fixing functional $F$ in the BRST-antiBRST
quantization \cite{BLT1, BLT2}.

The Yang--Mills theories belong to the class of irreducible gauge theories of
rank $1$ with a closed algebra, which implies that $M_{\alpha\beta}^{ij}=0 $
in (\ref{gauge_alg}) and that any solution of the equation $R_{\alpha}%
^{i}X^{\alpha}=0$ has the form $X^{\alpha}=0$. The corresponding space of
fields and antifields $\left(  \phi^{A},\phi_{Aa}^{\ast},\bar{\phi}\right)  $
is given by
\begin{equation}
\phi^{A}=\left(  A^{i},B^{\alpha},C^{\alpha a}\right)  \ ,\ \ \ \phi
_{Aa}^{\ast}=\left(  A_{ia}^{\ast},B_{\alpha a}^{\ast},C_{\alpha ab}^{\ast
}\right)  \ ,\ \ \ \bar{\phi}=\left(  \bar{A}_{i},\bar{B}_{\alpha},\bar
{C}_{aa}\right)  \ , \label{confSp}%
\end{equation}
as we take into account (\ref{antif}) and the following distribution of the
Grassmann parity and ghost number:%
\begin{equation}
\varepsilon(\phi^{A})\equiv\left(  \varepsilon_{i},\varepsilon_{\alpha
},\varepsilon_{\alpha}+1\right)  \ ,\ \ \ \mathrm{gh}(\phi^{A})=\left(
0,0,\left(  -1\right)  ^{a+1}\right)  \ , \label{Grpgh}%
\end{equation}
whereas a solution to the generating equations (\ref{3.3}) with a vanishing
right-hand side can be found in the linear form (\ref{linS}), $S=S_{0}%
+\phi_{Aa}^{\ast}X^{Aa}+\bar{\phi}_{A}Y^{A}$, obviously satisfying the
boundary condition $\left.  S\right\vert _{\phi^{\ast}=\bar{\phi}=0}=S_{0} $.
Here, the functionals $X^{Aa}$ and $Y^{A}$ can be chosen as \cite{BLT1}%
\begin{equation}
X^{Aa}=\left(  X_{1}^{ia},X_{2}^{\alpha a},X_{3}^{\alpha ab}\right)
\ ,\ \ \ Y^{A}=\left(  Y_{1}^{i},Y_{2}^{\alpha},Y_{3}^{\alpha a}\right)  \ ,
\label{solxy}%
\end{equation}
where%
\begin{align}
&  X_{1}^{ia}=R_{\alpha}^{i}C^{\alpha a}\ , &  &  X_{2}^{\alpha a}=-\frac
{1}{2}F_{\gamma\beta}^{\alpha}B^{\beta}C^{\gamma a}-\frac{1}{12}\left(
-1\right)  ^{\varepsilon_{\beta}}\left(  2F_{\gamma\beta,j}^{\alpha}R_{\rho
}^{j}+F_{\gamma\sigma}^{\alpha}F_{\beta\rho}^{\sigma}\right)  C^{\rho
b}C^{\beta a}C^{\gamma c}\varepsilon_{cb}\ ,\nonumber\\
&  X_{3}^{\alpha ab}=-\varepsilon^{ab}B^{\alpha}-\frac{1}{2}\left(  -1\right)
^{\varepsilon_{\beta}}F_{\beta\gamma}^{\alpha}C^{\gamma b}C^{\beta a}\ , &  &
Y_{1}^{i}=R_{\alpha}^{i}B^{\alpha}+\frac{1}{2}\left(  -1\right)
^{\varepsilon_{\alpha}}R_{\alpha,j}^{i}R_{\beta}^{j}C^{\beta b}C^{\alpha
a}\varepsilon_{ab}\ ,\nonumber\\
&  Y_{2}^{\alpha}=0\ , &  &  Y_{3}^{\alpha a}=-2X_{3}^{\alpha a}\ . \label{xy}%
\end{align}
By construction, the functionals $X^{Aa}=\delta S/\delta\phi_{Aa}^{\ast}$ and
$Y^{A}=\delta S/\delta\bar{\phi}_{A}$ obey the properties $S_{0,i}X^{ia}%
=0$,$\ X_{,B}^{Aa}X^{Bb}=\varepsilon^{ab}Y^{A}$,$\ Y_{,A}^{B}X^{Aa}=0$.
Besides, in Yang--Mills theories the explicit form (\ref{R(A)}), (\ref{F(A)})
of the gauge generators $R_{\alpha}^{i}$ and structure coefficients
$F_{\alpha\beta}^{\gamma}=\mathrm{const}$ is such that $X^{Aa}=\left(
X_{1}^{ia},X_{2}^{\alpha a},X_{3}^{\alpha ab}\right)  $ in (\ref{xy}) possess
the properties $X_{,A}^{Aa}=0$, so that the entire set of relations
(\ref{xy2}) is fulfilled, and the solution given by (\ref{xy}) satisfies the
generating equations (\ref{3.3}) identically.

As we keep the following consideration restricted to the case of constant
structure coefficients, $F_{\beta\gamma,j}^{\alpha}=0$, let us choose the
gauge-fixing functional $F\left(  \phi\right)  $ in the form%
\begin{equation}
F=F\left(  A,C\right)  \ ,\ \ \ \frac{\delta^{2}F}{\delta A^{i}\delta A^{j}%
}\neq0\ ,\ \ \ \frac{\delta^{2}F}{\delta C^{\alpha a}\delta C^{\alpha a}}%
\neq0\ . \label{gengF}%
\end{equation}
By virtue of (\ref{xy}), the quantum action $S_{F}(\phi)$ in (\ref{action})
reads as follows:%
\begin{align}
\ S_{F}  &  =S_{0}+\frac{\delta F}{\delta A^{i}}\left(  R_{\alpha}%
^{i}B^{\alpha}+\frac{1}{2}\left(  -1\right)  ^{\varepsilon_{\alpha}}%
R_{\alpha,j}^{i}R_{\beta}^{j}C^{\beta b}C^{\alpha a}\varepsilon_{ab}\right)
-\frac{1}{2}\varepsilon_{ab}\left(  R_{\alpha}^{i}C^{\alpha a}\right)
{\frac{\delta^{2}F}{\delta A^{i}\delta A^{j}}}\left(  R_{\beta}^{j}C^{\beta
b}\right) \nonumber\\
&  +\frac{\delta F}{\delta C^{\alpha a}}\left(  F_{\gamma\beta}^{\alpha
}B^{\beta}C^{\gamma a}+\frac{1}{6}\left(  -1\right)  ^{\varepsilon_{\beta}%
}F_{\gamma\sigma}^{\alpha}F_{\beta\rho}^{\sigma}C^{\rho b}C^{\beta a}C^{\gamma
c}\varepsilon_{cb}\right) \nonumber\\
&  -\frac{1}{2}\varepsilon_{ab}\left(  \varepsilon^{ac}B^{\alpha}+\frac{1}%
{2}\left(  -1\right)  ^{\varepsilon_{\gamma}}F_{\gamma\delta}^{\alpha
}C^{\delta c}C^{\gamma a}\right)  {\frac{\delta^{2}F}{\delta C^{\alpha
c}\delta C^{\beta d}}}\left(  \varepsilon^{bd}B^{\beta}+\frac{1}{2}\left(
-1\right)  ^{\varepsilon_{\rho}}F_{\rho\sigma}^{\beta}C^{\sigma d}C^{\rho
b}\right)  \ . \label{YMactg}%
\end{align}
Using the identity
\begin{align}
&  \frac{\delta F}{\delta A^{i}}R_{\alpha}^{i}B^{\alpha}+\frac{1}{2}\left(
-1\right)  ^{\varepsilon_{\alpha}}\varepsilon_{ab}\frac{\delta F}{\delta
A^{i}}R_{\alpha,j}^{i}R_{\beta}^{j}C^{\beta b}C^{\alpha a}-\frac{1}%
{2}\varepsilon_{ab}\left(  R_{\alpha}^{i}C^{\alpha a}\right)  {\frac
{\delta^{2}F}{\delta A^{i}\delta A^{j}}}\left(  R_{\beta}^{j}C^{b}\right)
\nonumber\\
&  \ =\chi_{\alpha}B^{\alpha}+\frac{1}{2}\left(  -1\right)  ^{\varepsilon
_{\alpha}}\left(  \chi_{\alpha,i}R_{\beta}^{i}\right)  C^{\beta b}C^{\alpha
a}\varepsilon_{ab}\,,\,\,\,\mathrm{for}\,\,\,\chi_{\alpha}\equiv\frac{\delta
F}{\delta A^{i}}R_{\alpha}^{i}\ , \label{ident}%
\end{align}
we obtain%
\begin{equation}
S_{F}=S_{0}+\frac{\delta F}{\delta A^{i}}\mathcal{A}^{i}-\frac{1}%
{2}\varepsilon_{ab}\left[  \frac{\delta}{\delta A^{j}}\left(  \frac{\delta
F}{\delta A^{i}}\mathcal{A}^{ia}\right)  \right]  \mathcal{A}^{jb}%
+\frac{\delta F}{\delta C^{\alpha a}}\mathcal{C}^{\alpha a}-\frac{1}%
{2}\varepsilon_{ab}\mathcal{C}^{\alpha ac}\left(  \frac{\delta}{\delta
C^{\beta d}}{\frac{\delta F}{\delta C^{\alpha c}}}\right)  \mathcal{C}^{\beta
bd}\ , \label{conv2}%
\end{equation}
where%
\begin{align}
&  \mathcal{A}^{i}\equiv R_{\alpha}^{i}B^{\alpha}\ ,\ \ \ \mathcal{A}%
^{ia}\equiv R_{\alpha}^{i}C^{\alpha a}\ ,\ \ \ \mathcal{C}^{\alpha a}\equiv
F_{\gamma\beta}^{\alpha}B^{\beta}C^{\gamma a}+\frac{1}{6}\left(  -1\right)
^{\varepsilon_{\beta}}F_{\gamma\sigma}^{\alpha}\left(  F_{\beta\rho}^{\sigma
}C^{\rho b}C^{\beta a}\right)  C^{\gamma c}\varepsilon_{cb}\ ,\label{note2}\\
&  \ \mathcal{C}^{\alpha ab}\equiv\varepsilon^{ab}B^{\alpha}+\frac{1}%
{2}\left(  -1\right)  ^{\varepsilon_{\beta}}F_{\beta\gamma}^{\alpha}C^{\gamma
b}C^{\beta a}\ ,\ \ \ \mathrm{with}\ \ \ \varepsilon\left(  \mathcal{A}%
^{i}\right)  =\varepsilon\left(  \mathcal{A}^{ia}\right)  +1=\varepsilon
_{i}\ ,\ \ \ \varepsilon\left(  \mathcal{C}^{\alpha ab}\right)  =\varepsilon
\left(  \mathcal{C}^{\alpha a}\right)  +1=\varepsilon_{\alpha}\ .\nonumber
\end{align}

For Yang--Mills theories, with the classical action $S_{0}$, gauge generators
$R_{\alpha}^{i}$ and structure coefficients $F_{\alpha\beta}^{\gamma}$ given
by (\ref{4.1}), (\ref{R(A)}), (\ref{F(A)}), and with the set of fields
$\phi^{A}\,$given by (\ref{YM-fields}), (\ref{YM-e,gh}), the relations
(\ref{conv2}), (\ref{note2}) take the form%
\begin{align}
S_{F}  &  =S_{0}+\int d^{D}x\ \left\{  \frac{\delta F}{\delta A^{m\mu}%
}\mathcal{A}^{m\mu}-\frac{1}{2}\varepsilon_{ab}\left[  \frac{\delta}{\delta
A^{n\nu}}\left(  \frac{\delta F}{\delta A^{m\mu}}\mathcal{A}^{m\mu a}\right)
\mathcal{A}^{n\nu b}\right]  \right\} \nonumber\\
&  +\int d^{D}x\left[  \frac{\delta F}{\delta C^{ma}}\mathcal{C}^{ma}-\frac
{1}{2}\varepsilon_{ab}\mathcal{C}^{mac}\left(  \frac{\delta}{\delta C^{nd}%
}{\frac{\delta F}{\delta C^{mc}}}\right)  \mathcal{C}^{nbd}\right]  \ ,
\label{conv2y}%
\end{align}
where%
\begin{align}
&  \mathcal{A}_{\mu}^{m}\equiv D_{\mu}^{mn}B^{n}\ ,\ \ \ \mathcal{A}_{\mu
}^{ma}\equiv D_{\mu}^{mn}C^{na}\ ,\ \ \ \mathcal{C}^{ma}\equiv f{^{mnl}%
B^{l}C^{na}}+\frac{1}{6}f{^{mnl}}\left(  f^{lrs}C^{sb}C^{ra}\right)
C^{nc}\varepsilon_{cb}\ ,\label{note2y}\\
&  \mathcal{C}^{mab}\equiv\varepsilon^{ab}B^{m}+\frac{1}{2}f^{mnl}C^{lb}%
C^{na}\ ,\ \ \ \varepsilon\left(  \mathcal{A}_{\mu}^{m}\right)  =\varepsilon
\left(  \mathcal{A}_{\mu}^{ma}\right)  +1=0\ ,\ \ \ \varepsilon\left(
\mathcal{C}^{ma}\right)  =\varepsilon\left(  \mathcal{C}^{mab}\right)
+1=1\ .\nonumber
\end{align}
Choosing the gauge-fixing functional $F\left(  A,C\right)  $ in the quadratic
form (\ref{F(A,C)}) and using the identities (for arbitrary $su(N)$-vectors
$F^{m}$ and $G^{m}$)%
\begin{equation}
D_{\mu}^{mn}A^{n\mu}=\partial_{\mu}A^{m\mu}\ ,\ \ \ \int d^{D}x\ \left(
D_{\mu}^{mn}F^{n}\right)  G^{m}=-\int d^{D}x\ F^{m}D_{\mu}^{mn}G^{n}\ ,
\label{idDmu}%
\end{equation}
we have%
\begin{align}
&  \delta_{A}F=-\alpha\int d^{D}x\ A_{\mu}^{m}\delta A^{m\mu}\ ,\\
&  \frac{\delta F}{\delta A^{m\mu}}\mathcal{A}^{m\mu}=-\alpha\int
d^{D}x\ A^{m\mu}D_{\mu}^{mn}B^{n}=\alpha\int d^{D}x\ \left(  D_{\mu}%
^{nm}A^{m\mu}\right)  B^{n}=\alpha\int d^{D}x\ \left(  \partial_{\mu}A^{m\mu
}\right)  B^{mn}\ ,\\
&  \frac{\delta F}{\delta A^{m\mu}}\mathcal{A}^{m\mu a}=-\alpha\int
d^{D}x\ A^{m\mu}D_{\mu}^{mn}C^{na}=\alpha\int d^{D}x\ \left(  \partial_{\mu
}A^{n\mu}\right)  C^{na}\ ,
\end{align}
whence%
\begin{align}
&  \ \delta_{A}\left(  \frac{\delta F}{\delta A^{m\mu}}\mathcal{A}^{m\mu
a}\right)  =\alpha\int d^{D}x\ \left(  \partial_{\mu}\delta A^{m\mu}\right)
C^{ma}=-\alpha\int d^{D}x\ \left(  \partial_{\mu}C^{ma}\right)  \delta
A^{m\mu}\ ,\nonumber\\
&  \int d^{D}x\ \left[  \frac{\delta}{\delta A^{n\nu}}\left(  \frac{\delta
F}{\delta A^{m\mu}}\mathcal{A}^{m\mu a}\right)  \right]  \mathcal{A}^{n\nu
b}=-\alpha\int d^{D}x\ \left(  \partial_{\mu}C^{ma}\right)  D^{mn\mu}C^{nb}\ .
\end{align}
Next,%
\begin{align}
&  \delta_{C}F=-\beta\varepsilon_{ba}\int d^{D}x\ C^{mb}\delta C^{ma}%
\ \Longrightarrow\ \frac{\delta F}{\delta C^{ma}}=\beta\varepsilon_{ab}%
C^{mb}\ ,\\
&  \int d^{D}x\frac{\delta F}{\delta C^{ma}}\mathcal{C}^{ma}=\beta
\varepsilon_{ab}\int d^{D}x\ C^{mb}\mathcal{C}^{ma}=\beta\varepsilon_{ba}\int
d^{D}x\ C^{ma}\left(  f{^{mnl}B^{l}C^{nb}}+\frac{1}{6}f{^{mnl}}f^{lrs}%
C^{sd}C^{rb}C^{nc}\varepsilon_{cd}\right)  \ .
\end{align}
At the same time,%
\begin{align}
&  \ \delta_{C}\left(  \frac{\delta F}{\delta C^{mc}\left(  x\right)
}\right)  =\beta\varepsilon_{cd}\delta C^{md}\left(  x\right)  =\beta
\varepsilon_{cd}\int d^{D}y\ \delta^{mn}\delta\left(  y-x\right)  \delta
C^{nd}\left(  y\right)  ~,\nonumber\\
&  \ \frac{\delta}{\delta C^{nd}\left(  y\right)  }\left(  \frac{\delta
F}{\delta C^{mc}\left(  x\right)  }\right)  =\beta\varepsilon_{cd}\delta
^{mn}\delta\left(  y-x\right)  \ ,
\end{align}
whence%
\begin{align}
&  -\frac{1}{2}\varepsilon_{ab}\int d^{D}x\ d^{D}y\ \mathcal{C}^{mac}\left(
x\right)  \frac{\delta}{\delta C^{nd}\left(  y\right)  }\left(  {\frac{\delta
F}{\delta C^{mc}\left(  x\right)  }}\right)  \mathcal{C}^{nbd}\left(  y\right)
\nonumber\\
&  \ \ =-\frac{1}{2}\varepsilon_{ab}\int d^{D}x\ d^{D}y\ \mathcal{C}%
^{mac}\left(  x\right)  \left[  \beta\varepsilon_{cd}\delta^{mn}\delta\left(
y-x\right)  \right]  \mathcal{C}^{nbd}\left(  y\right) \\
&  \ \ =-\frac{\beta}{2}\varepsilon_{ab}\varepsilon_{cd}\int d^{D}x\ \left(
\varepsilon^{ac}B^{m}+\frac{1}{2}f^{mnl}C^{lc}C^{na}\right)  \left(
\varepsilon^{bd}B^{m}+\frac{1}{2}f^{mrs}C^{sd}C^{rb}\right)  \ .\nonumber
\end{align}
Therefore,%
\begin{align}
&  \hspace{-1em}\ \int d^{D}x\left[  \frac{\delta F}{\delta C^{ma}}%
\mathcal{C}^{ma}-\frac{1}{2}\varepsilon_{ab}\mathcal{C}^{mac}\frac{\delta
}{\delta C^{nd}}\left(  {\frac{\delta F}{\delta C^{mc}}}\right)
\mathcal{C}^{nbd}\right] \nonumber\\
&  =-\beta\varepsilon_{ab}\int d^{D}x\ C^{ma}\left(  f{^{mnl}B^{l}C^{mb}%
}+\frac{1}{6}f{^{mnl}}f^{lrs}C^{sd}C^{rb}C^{nc}\varepsilon_{cd}\right)
\label{aux1}\\
&  \ -\frac{\beta}{2}\varepsilon_{ab}\varepsilon_{cd}\int d^{D}x\ \left(
\varepsilon^{ac}B^{m}+\frac{1}{2}f^{mnl}C^{lc}C^{na}\right)  \left(
\varepsilon^{bd}B^{m}+\frac{1}{2}f^{mrs}C^{sd}C^{rb}\right)  \ .\nonumber
\end{align}
Finally,%
\begin{equation}
S_{F}(A,B,C)=S_{0}\left(  A\right)  +S_{1}\left(  A,B\right)  +S_{2}\left(
A,C\right)  +S_{3}(A,B,C)\ , \label{S(A,B,C)2}%
\end{equation}
where%
\begin{align}
S_{1}  &  =\alpha\int d^{D}x\ \left(  \partial^{\mu}A_{\mu}^{m}\right)
B^{m}\ ,\quad S_{2}=\frac{\alpha}{2}\varepsilon_{ab}\int d^{D}x\ \left(
\partial^{\mu}C^{ma}\right)  D_{\mu}^{mn}C^{nb}\ ,\nonumber\\
S_{3}  &  =-\beta\varepsilon_{ab}\int d^{D}x\ C^{ma}\left(  f{^{mnl}%
B^{l}C^{mb}}+\frac{1}{6}f{^{mnl}}f^{lrs}C^{sd}C^{rb}C^{nc}\varepsilon
_{cd}\right) \nonumber\\
&  -\frac{\beta}{2}\varepsilon_{ab}\varepsilon_{cd}\int d^{D}x\ \left(
\varepsilon^{ac}B^{m}+\frac{1}{2}f^{mnl}C^{lc}C^{na}\right)  \left(
\varepsilon^{bd}B^{m}+\frac{1}{2}f^{mrs}C^{sd}C^{rb}\right)  \ .
\end{align}
By virtue of the identity $f{^{lmn}C^{nb}}C^{ma}\varepsilon_{ab}\equiv0$, the
quantum action (\ref{S(A,B,C)2}) equals to (\ref{S(A,B,C)}).

\end{document}